\documentclass[final]{ws-ijmpe}
\usepackage[super,compress]{cite}
\usepackage{units}
\usepackage{xspace}
\usepackage{subfig}
\usepackage{hyperref}
\def\draftnote{}

\newcommand{\vect}[1]{\vec{\mathbf{#1}}}
\newcommand{\vectS}[1]{\vec{\boldsymbol{#1}}}

\newcommand{\EFT}{$\mathrm{EFT}_{\not{\pi}}$\xspace}

\newcommand{\srutTypeOne}[1]{\vrule width0pt height0pt depth #1\relax}
\newcommand{\jjvHe}{{}^3\mathrm{He}}
\newcommand{\jjvH}{{}^3\mathrm{H}}

\begin{document}

\markboth{Jared Vanasse}{Three-Body Systems In Pionless Effective Field Theory}

\catchline{}{}{}{}{}

\title{Three-body systems in pionless effective field theory}

\author{Jared Vanasse}

\address{Department of Physics, Duke University, \\
Durham, NC 27708,
USA\\
jjv9@phy.duke.edu}

\maketitle

\begin{history}
\received{Day Month Year}
\revised{Day Month Year}
\end{history}

\begin{abstract}
Investigations of three-body nuclear systems using pionless effective field theory (\EFT) are reviewed.  The history of \EFT in $nd$ and $pd$ scattering is briefly discussed and emphasis put on the use of strict perturbative techniques.  In addition renormalization issues appearing in $pd$ scattering are also presented.  Bound state calculations are addressed and new perturbative techniques for describing them are highlighted.  Three-body breakup observables in $nd$ scattering are also considered and the utility of \EFT for addressing them.
\end{abstract}


\ccode{PACS numbers:25.10+s, 21.30.Fe, 21.45 Ff, 25.40.Dn}


\section{Introduction}

The essential ingredient for any effective field theory (EFT) is the power counting, which orders contributions in powers of $(Q/\Lambda_{b})^{n}$, where $Q$ is the typical momentum scale and $\Lambda_{b}$ the breakdown scale, where new physics not explicitly encoded in the effective description occurs.  The size of low energy constants (LECs) may be estimated by naive dimensional analysis (NDA), where the size of LECs is determined by the scales in the theory, typically $\Lambda_{b}$.  However, in two and three-body nuclear systems it is found that NDA is not adequate.  In the two-body system an unnatural scale corresponding to the deuteron binding momentum is created by the importance of a non-perturbative resummation that leads to a fine tuning between scales\cite{Bedaque:1997qi,vanKolck:1997ut,Kaplan:1998tg,Kaplan:1998we}.  Thus, for energies greater than the deuteron binding energy, $E_{d}^{B}=2.22$~MeV, leading-order (LO) interactions must be treated non-perturbatively to obtain a consistent power counting and reproduce the deuteron bound state.  For energies below the deuteron binding energy the individual nucleons in the deuteron cannot be resolved and it can be treated as a fundamental degree of freedom leading to a perturbative description of nuclear interactions.

For low energies ($E\lesssim m_{\pi}^{2}/M_{N}$) pions are not dynamical and a theory containing only nucleons and external currents as degrees of freedom is appropriate.  This theory, known as pionless effective field theory (\EFT), has been used to great effect in the two-body sector for calculating $NN$ scattering \cite{Chen:1999tn,Ando:2004mm,Ando:2007fh,Kong:1999sf} and deuteron electromagnetic form factors \cite{Chen:1999tn}.  It has also yielded a precision calculation ($<$1\%) of the $np$ capture process. \cite{Chen:1999bg,Rupak:1999rk,Ando:2005cz}.  Parity violating (PV) interactions have also been calculated \cite{Schindler:2009wd,Phillips:2008hn,Shin:2009hi,Vanasse:2014sva} as well as neutrino-deuteron processes \cite{Kong:2000px,Butler:2000zp,Ando:2008va,Chen:2012hm}.  \EFT has also been applied extensively in the three-body sector and that will be the focus of this work.  In the three-body sector most calculations have been done in momentum space with only a few \EFT calculations performed in configuration space.  For further details of configuration space techniques and results consult Refs.~\refcite{Kirscher:2009aj}, \refcite{Kirscher:2011uc}, and \refcite{Kirscher:2015ana}.  Here we will focus exclusively on momentum space calculations.

Section~\ref{sec:twobody} contains a brief review of two-body physics to the extent necessary to understand three-body systems.  Section~\ref{sec:3BScattering} will offer a brief review of the history of $nd$ scattering in \EFT.  In addition it will highlight the most recent numerical techniques in $nd$ scattering, and briefly address three-body forces.  Section~\ref{sec:boundstates} contains a review of recent advances in perturbative calculations for three-body bound states in \EFT.  Calculations of the triton charge radius, and a novel approach to calculating three-body forces will be discussed. Section~\ref{sec:Coulomb} will deal with the inclusion of Coulomb forces in $pd$ scattering and review recent findings of the need for a new isospin-dependent counterterm at NLO.  Finally, section \ref{sec:3BBreakup} briefly considers the possibility of using \EFT to probe three-body breakup observables. Conclusions are in Section~\ref{sec:conclusion}.

\section{\label{sec:twobody}Two-Body System}

The Lagrangian in the two-body sector of \EFT is
\begin{align}
\mathcal{L}_{2}=\ &\hat{N}^{\dagger}\left(i\partial_{0}+\frac{\vect{\nabla}^2}{2M_{N}}\right)\hat{N}+\hat{t}_{i}^{\dagger}\left(\Delta_{t}-c_{0t}\left(i\partial_{0}+\frac{\vect{\nabla}^{2}}{4M_{N}}+\frac{\gamma_{t}^{2}}{M_{N}}\right)\right)\hat{t}_{i}\\\nonumber
&+\hat{s}_{a}^{\dagger}\left(\Delta_{s}-c_{0s}\left(i\partial_{0}+\frac{\vect{\nabla}^{2}}{4M_{N}}+\frac{\gamma_{s}^{2}}{M_{N}}\right)\right)\hat{s}_{a}\\\nonumber
&+y_{t}\left[\hat{t}_{i}^{\dagger}\hat{N} ^{T}P_{i}\hat{N} +\mathrm{H.c.}\right]+y_{s}\left[\hat{s}_{a}^{\dagger}\hat{N}^{T}\bar{P}_{a}\hat{N}+\mathrm{H.c.}\right],
\end{align}
where the auxiliary field formalism is used, $\hat{N}$ is a nucleon field, and $\hat{t}_{i}$ ($\hat{s}_{a}$) is a deuteron (spin-singlet dibaryon) field, with $P_{i}=\frac{1}{\sqrt{8}}\sigma_{2}\sigma_{i}\tau_{2}$ ($\bar{P}_{a}=\frac{1}{\sqrt{8}}\sigma_{2}\tau_{2}\tau_{a}$) projecting out the spin-triplet iso-singlet (spin-singlet iso-triplet) channel.  The last line represents the two-body contact interactions.  In practice these parameters are fit using the effective range expansion (ERE) or the $Z$-parametrization\cite{Phillips:1999hh,Griesshammer:2004pe}.  Here the $Z$-parametrization is used where at LO the fit is to the deuteron bound state pole in the ${}^{3}\!S_{1}$ channel and the virtual bound state pole in the ${}^{1}\!S_{0}$ channel.  At NLO and N$^{2}$LO the parameters are fit to ensure the poles are at the same position and have the correct residues.  In the $Z$-parametrization the parameters are
\begin{align}
&y_{t}^{2}=\frac{4\pi}{M_{N}},\quad \Delta_{t}=\gamma_{t}-\mu,\quad c_{0t}^{(n)}=(-1)^{n}(Z_{t}-1)^{n+1}\frac{M_{N}}{2\gamma_{t}}\\\nonumber
&y_{s}^{2}=\frac{4\pi}{M_{N}},\quad \Delta_{s}=\gamma_{s}-\mu,\quad c_{0s}^{(n)}=(-1)^{n}(Z_{s}-1)^{n+1}\frac{M_{N}}{2\gamma_{s}},
\end{align}
where $\gamma_{t}=45.7025$~MeV ($\gamma_{s}=-7.890$~MeV) is the deuteron binding momentum (${}^{1}\!S_{0}$ virtual bound state pole binding momentum), $Z_{t}=1.6908$ ($Z_{s}=.9015$) is the residue about the deuteron pole (${}^{1}\!S_{0}$ virtual bound state pole), and $\mu$ a scale introduced by dimensional regularization with the power divergence subtraction (PDS) scheme\cite{Kaplan:1998tg,Kaplan:1998we}. \footnote{The scale $\mu$ in the LECs cancels with a scale $\mu$ from dimensionally regularized integrals with PDS such that the amplitude is independent of $\mu$.} Note that $c_{0s,t}$ gets corrections at each order beyond NLO.

\begin{figure}[hbt]
\hspace{.6cm}
\includegraphics[width=110mm]{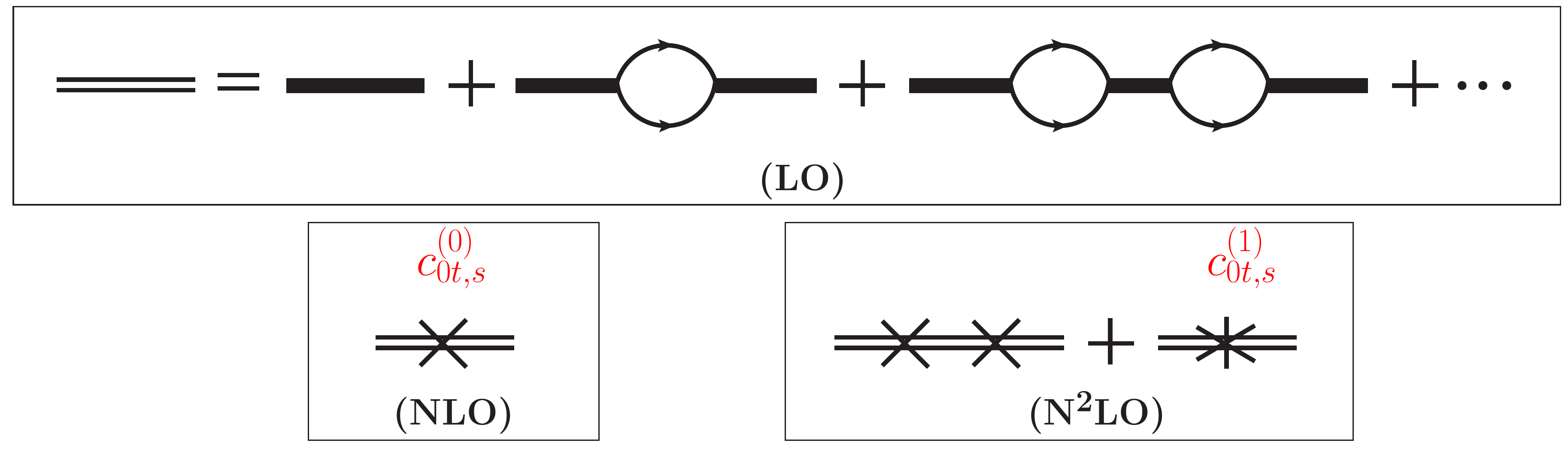}
\caption{\label{fig:DeutProp}  The bare dibaryon propagator, $i/\Delta_{t,s}$, is represented by a thick solid line, and nucleon propagators by thin lines.  The infinite sum of bubble diagrams gives the dressed dibaryon propagator represented by a double line.  The NLO correction to the dibaryon propagator has one insertion of $c_{0t,s}^{(0)}$ represented by a cross.  The N$^{2}$LO correction to the dibaryon propagator has two insertions of $c_{0t,s}^{(0)}$ and one of $c_{0t,s}^{(1)}$ represented by a star.}
\end{figure}

The LO dibaryon propagator is given by the bubble sum in Fig.~\ref{fig:DeutProp}
where the thick line is the bare dibaryon propagator, $i/\Delta_{s,t}$, and the thin lines with arrows nucleon propagators.  The NLO correction to the dibaryon propagator is given by a single effective range insertion, $c_{0t,s}^{(0)}$, which is represented by a cross.  At N$^{2}$LO the dibaryon propagator receives two insertions of $c_{0t,s}^{(0)}$ and one insertion of $c_{0t,s}^{(1)}$ as shown in Fig.~\ref{fig:DeutProp}.  The deuteron and spin-singlet dibaryon propagator up to and including N$^{2}$LO in the $Z$-parametrization are given by\cite{Griesshammer:2004pe}
\begin{align}
&iD_{\{t,s\}}^{\mathrm{N}^{2}\mathrm{LO}}(p_{0},\vect{p})=\frac{i}{\gamma_{\{t,s\}}-\sqrt{\frac{\vect{p}^{2}}{4}-M_{N}p_{0}-i\epsilon}}\times\\\nonumber
&\times\left[\underbrace{1\vphantom{\frac{Z_{\{t,s\}}-1}{2\gamma_{\{t,s\}}}\left(\gamma_{\{t,s\}}+\sqrt{\frac{\vect{p}^{2}}{4}-M_{N}p_{0}-i\epsilon}\right)}}_{\mathrm{LO}}+\underbrace{\frac{Z_{\{t,s\}}-1}{2\gamma_{\{t,s\}}}\left(\gamma_{\{t,s\}}+\sqrt{\frac{\vect{p}^{2}}{4}-M_{N}p_{0}-i\epsilon}\right)}_{\mathrm{NLO}}\right.\\\nonumber
&\left.\hspace{.5cm}+\underbrace{\left(\frac{Z_{\{t,s\}}-1}{2\gamma_{\{t,s\}}}\right)^{2}\left(\frac{\vect{p}^{2}}{4}-M_{N}p_{0}-\gamma_{\{t,s\}}^{2}\right)}_{\mathrm{NNLO}}+\cdots\right]\\\nonumber
\end{align}
From the residue of the deuteron propagator the deuteron wavefunction renormalization is given by
\begin{equation}
\label{eq:ZD}
Z_{D}=\frac{2\gamma_{t}}{M_{N}}\left[\underbrace{\srutTypeOne{.1cm}1}_{\mathrm{LO}}+\underbrace{(Z_{t}-1)}_{\mathrm{NLO}}+\underbrace{\srutTypeOne{.1cm}0}_{\mathrm{NNLO}}+\cdots\right].
\end{equation}
By construction the deuteron residue is reproduced exactly at NLO in the $Z$-parametrization.  The LO deuteron wavefunction renormalization will be defined by
\begin{equation}
Z_{\mathrm{LO}}=\frac{2\gamma_{t}}{M_{N}}.
\end{equation}

\section{\label{sec:3BScattering} Three-Body Scattering}
\subsection{Introduction}

The first three-body calculations in \EFT were carried out for $nd$ scattering in the quartet channel ($S=\nicefrac{3}{2}$) as it is qualitatively simpler than the doublet channel ($S=\nicefrac{1}{2}$).  Bedaque and van Kolck calculated the LO quartet $S$-wave channel scattering length, in which they resummed the effective range\cite{Bedaque:1997qi}.  Shortly thereafter with Hammer they considered the energy dependence in the quartet $S$-wave channel again with a resummed effective range \cite{Bedaque:1998mb}.  Then in the doublet $S$-wave channel they showed a three-body force at LO is required to properly renormalize results\cite{Bedaque:1998kg,Bedaque:1998km,Bedaque:1999ve,Hammer:2000nf}.  With this new non-perturbative renormalization they predicted the energy dependence at LO in the doublet $S$-wave channel\cite{Bedaque:1999ve}.

NLO calculations were then carried out in the quartet $S$-wave channel by Bedaque and Grie{\ss}hammer \cite{Bedaque:1999vb} and in the doublet $S$-wave channel by Hammer and Mehen \cite{Hammer:2001gh}.  Higher partial waves (up to and including $G$-waves) with the exception of the doublet $S$-wave were then calculated to N$^{2}$LO by Gabbiani et al.\cite{Gabbiani:1999yv}. However, at N$^{2}$LO they used the dibaryon propagator with fully resummed range corrections, and therefore their calculation was not strictly perturbative at N$^{2}$LO, as it included range corrections to all orders.  The doublet $S$-wave channel was finally addressed at N$^{2}$LO by Bedaque et al\cite{Bedaque:2002yg}. In this work they introduced the partial resummation technique for calculating higher order contributions and showed that a new energy dependent three-body force is required at N$^{2}$LO.  However, the partial resummation technique again suffered from not being strictly perturbative.  It was later shown by Platter and Phillips for cold atom calculations that if the cutoff is taken to infinity that the N$^{2}$LO energy dependent three-body force is not needed in the partial resummation technique\cite{Platter:2006ev}.  However, Ji and Phillips showed in cold atom systems that in a strictly perturbative calculation that a N$^{2}$LO energy dependent three-body force is required~\cite{Ji:2012nj}.  Calculations of all partial waves using the partial resummation technique were later revisited to N$^{2}$LO by Grie{\ss}hammer using the $Z$-parametrization in order to improve convergence to physical results\cite{Griesshammer:2004pe}.  Separately Gabbiani \cite{Gabbiani:2001yh} and Grie{\ss}hammer \cite{Griesshammer:2004pe} considered the use of fully resummed range corrections in dibaryon propagators with differing results. Grie{\ss}hammer found a three-body force was still needed in the doublet $S$-wave channel while Gabbiani did not.  Despite differing results both considered the use of fully resummed range corrections to be problematic in practical applications.

Formal investigations of the power counting of three-body forces, using naive dimensional analysis, were carried out by Grie{\ss}hammer \cite{Griesshammer:2005ga} and Birse \cite{Birse:2005pm}, and for the PV sector by Grie{\ss}hammer and Schindler \cite{Griesshammer:2010nd}.  Calculation of the PV spin rotation of a neutron through deuterium were carried out separately by Vanasse at LO \cite{Vanasse:2011nd} and Grie{\ss}hammer et al.\cite{Griesshammer:2011md} at NLO using the partial resummation technique and $Z$-parametrization.  The calculation of $nd$ scattering was then improved by Vanasse \cite{Vanasse:2013sda} in which a technique to calculate higher order corrections strictly perturbatively was developed.  In addition he considered two-body $SD$ mixing, which allowed for the investigation of polarization observables in $nd$ scattering.  However, at this order poor agreement was found with available data and potential model calculations (PMC).  This work was then improved by Vanasse with the perturbative technique being slightly improved and extended to bound states\cite{Vanasse:2015fph}, building upon the LO calculation of Hagen et al \cite{Hagen:2013xga} in halo EFT by calculating higher order contributions in \EFT.  Hammer and K{\"o}nig investigated the possibility of bound di-neutrons by calculating the dependence of three-body observables on the $nn$ scattering length\cite{Hammer:2014rba}.  Finally, Margaryan et al.~calculated polarization observables in $nd$ scattering to N$^{3}$LO in \EFT by considering contributions from two-body $P$-wave contact interactions \cite{Margaryan:2015rzg}.

Below the formalism for $nd$ scattering is introduced starting with the quartet channel, and then proceeded by the doublet channel.  In addition the partial resummation technique is briefly reviewed but the focus is on the newer strictly perturbative techniques.  Three-body forces will briefly be discussed, while a different approach will be addressed in a later section.

\subsection{Quartet Channel}

\begin{figure}[th]
\centerline{\includegraphics[width=100mm]{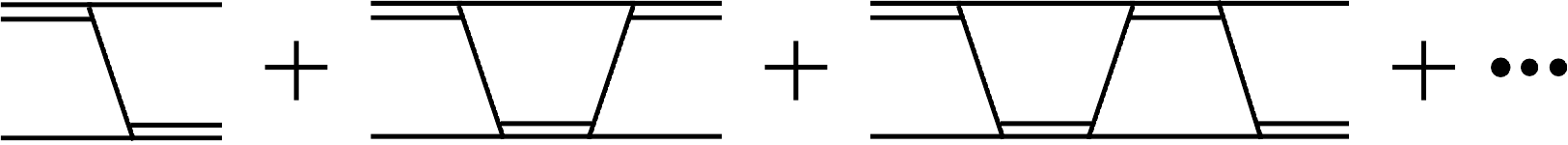}}
\caption{\label{fig:LODiagrams}Infinite sum of diagrams contributing to LO quartet channel $nd$ scattering amplitude.  The double line represents a dressed deuteron propagator and the single line a nucleon propagator.}
\end{figure}

\begin{figure}[th]
\centerline{\includegraphics[width=100mm]{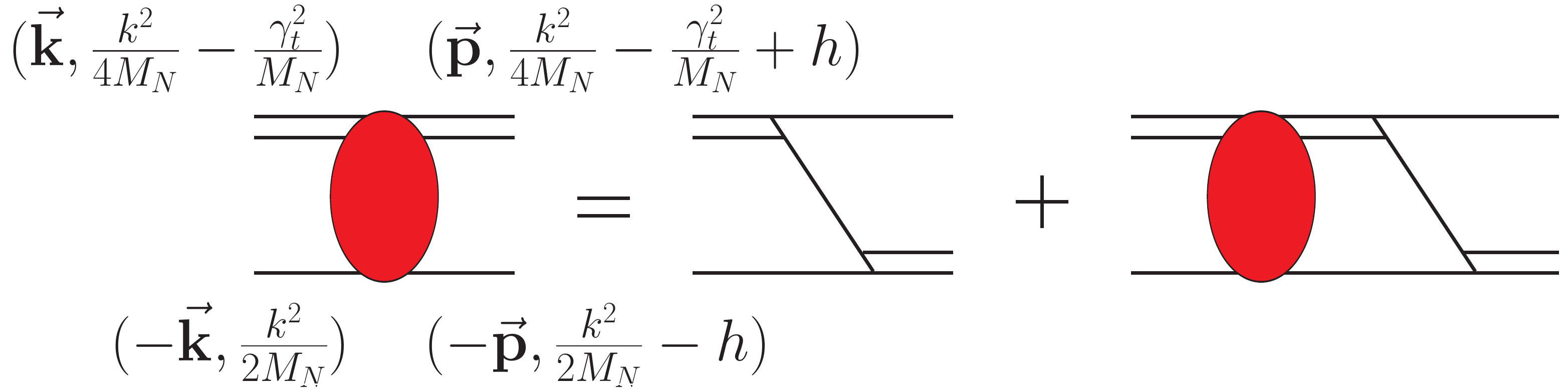}}
\caption{\label{fig:LOIntegralEquation} Integral equation for LO $nd$ scattering amplitude in quartet channel.  The momentum $\vect{k}$ ($\vect{p}$) is the incoming (outgoing) momentum in the center of mass frame that is on-shell (off-shell).  The parameter $h$ is the off shell parameter for $\vect{p}$.  When $h=0$ then $|\vect{k}|=|\vect{p}|$.}
\end{figure}

At LO in \EFT $nd$ scattering in the quartet channel is given by the infinite sum of diagrams represented in Fig.~\ref{fig:LODiagrams}.  At LO these diagrams all scale as
$\Lambda_{\not{\pi}\xspace}/(M_{N}Q^{2})$, where $\Lambda_{\not{\pi}\xspace}\sim m_{\pi}$\cite{Bedaque:1999vb}. \footnote{Note that if properly renormalized with the deuteron wavefunction renormalization they scale as $1/(M_{N}Q)$, exactly as in the LO two-body case.}
Unfortunately, the explicit sum of these diagrams seems to offer no immediate analytical solution as in the two-body case, rather this sum of diagrams is rewritten as an integral equation given in Fig.~\ref{fig:LOIntegralEquation}.
Projecting the integral equation onto the quartet channel and a partial wave basis gives\cite{Skornyakov,Gabbiani:1999yv}
\begin{align} 
\label{eq:LOndScattQuartet}
t^{\ell}_{0,q}(k,p)=&-\frac{y_{t}^{2}M_{N}}{pk}Q_{\ell}\left(\frac{p^{2}+k^{2}-M_{N}E-i\epsilon}{pk}\right)\\\nonumber
&-\frac{2}{\pi}\int_{0}^{\Lambda}dq q^{2}t_{0.q}^{\ell}(k,q)\frac{1}{\sqrt{\frac{3}{4}q^{2}-M_{N}E-i\epsilon}-\gamma_{t}}\frac{1}{qp}Q_{\ell}\left(\frac{p^{2}+q^{2}-M_{N}E-i\epsilon}{pq}\right),
\end{align}
where $Q_{\ell}(a)$ are Legendre functions of the second kind defined by\footnote{Note the convention used here for Legendre functions of the second kind differs from the standard convention by a phase of $(-1)^{\ell}$.}
\begin{equation}
Q_{\ell}(a)=\frac{1}{2}\int_{-1}^{1}dx\frac{P_{\ell}(x)}{x+a},
\end{equation}
with $P_{\ell}(x)$ being the standard Legendre polynomials.  The incoming momentum $\vect{k}$ in the center of mass (c.m.) frame is on shell and thus $M_{N}E=\frac{3}{4}k^{2}-\gamma_{t}^{2}$, while the outgoing momentum $\vect{p}$ is off shell. The parameter $h$ in Fig.~\ref{fig:LOIntegralEquation} is the off-shell parameter.  In Eq. (\ref{eq:LOndScattQuartet}) we set $h=(p^{2}-k^{2})/2M_{N}$\cite{}\cite{Bedaque:1999vb} to put the outgoing nucleon leg on-shell but keep the outgoing deuteron leg off-shell.  This is useful for three-body breakup.  The typical method for solving this integral equation is the Nystrom method\cite{press1996numerical}.  When $-\gamma_{t}^{2}<M_{N}E<0$ the only singularity in the integral equation is due to the deuteron pole.  This singularity is fixed and can be addressed by using a principal value prescription and standard subtraction techniques\cite{delves1988computational}.  The integral equation can also be rewritten using the $K$-matrix\cite{glockle2012quantum}, which has the advantage of avoiding the use of complex numbers in computations.  However, when $M_{N}E>0$ branch point singularities are encountered due to the three-body breakup channel.  The location of these singularities is not fixed and therefore cannot be dealt with by simple subtraction procedures.  However, these singularities are logarithmic and can be integrated over.  Thus choosing a large number of mesh points can reduce the numerical noise from these singularities and approach the actual solution.  Another approach is to avoid these singularities by rotating the path of integration into the complex plane.  Once the amplitude is solved on this contour the integral equation can be used again to rotate the solution back to the real axis.  This method, the Hetherington-Schick method \cite{Hetherington:1965zza,Ziegelmann}, has been used to great success in calculating these integral equations and has been put on firm mathematical grounding\cite{Brayshaw:1969ab}.

\begin{figure}[th]
\centerline{\includegraphics[width=30mm]{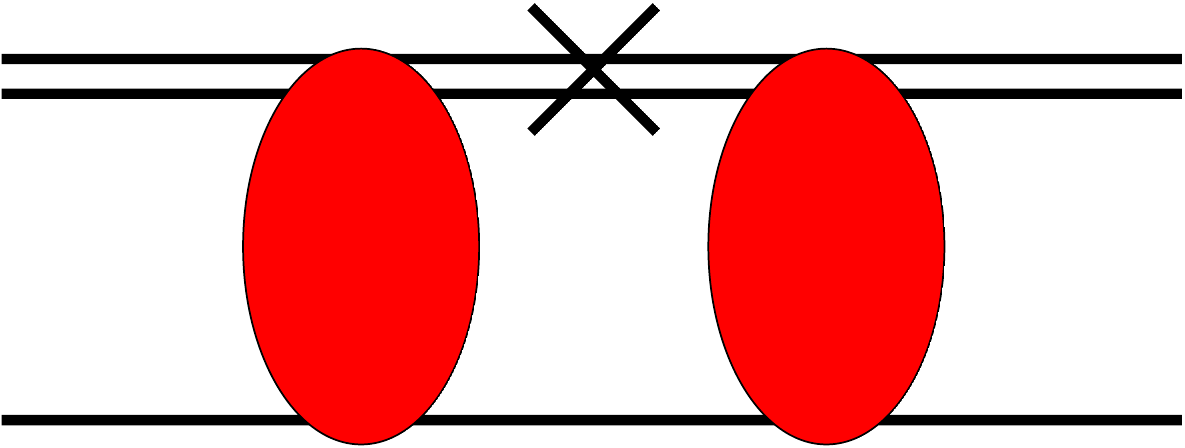}}
\caption{\label{fig:NLOCorrectionDiagram} NLO correction to $nd$ scattering amplitude in quartet channel.}
\end{figure}

\begin{figure}[th]
\centerline{\includegraphics[width=75mm]{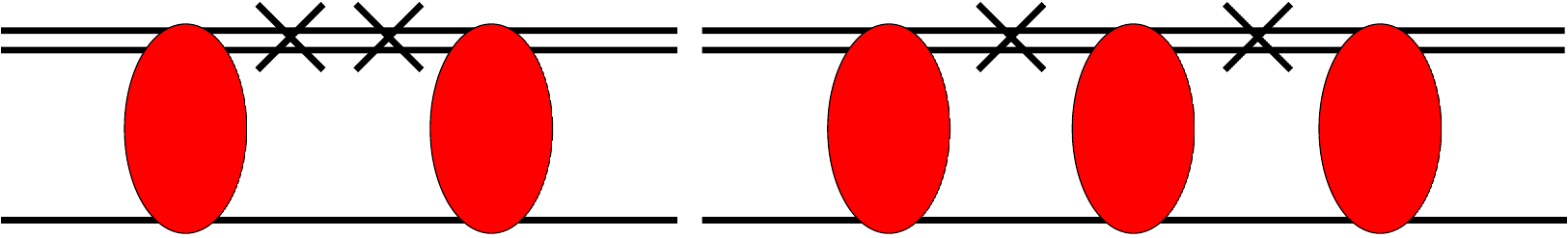}}
\caption{\label{fig:NNLOCorrectionDiagram}N$^{2}$LO correction to $nd$ scattering amplitude in quartet channel.}
\end{figure}

According to the power counting of \EFT in the $Z$-parametrization the LO solution will roughly require corrections of 35\% ($[Z_{t}-1]/2\approx.35$).  The NLO correction to the $nd$ scattering amplitude is given by the diagram in Fig.~\ref{fig:NLOCorrectionDiagram}, where the cross represents an effective range insertion.  This diagram contains two half off-shell LO $nd$ scattering amplitudes that can be numerically integrated to yield the NLO correction.   The N$^{2}$LO correction to the $nd$ scattering amplitude is given by the diagrams in Fig.~\ref{fig:NNLOCorrectionDiagram}. 
In the second diagram we see there is a full off-shell\footnote{For the full off-shell scattering amplitude both the incoming momentum $k$, and outgoing momentum $p$ do not satisfy the on-shell condition $M_{N}E=\frac{3k^{2}}{4M_{N}}-\gamma_{t}^{2}$.} $nd$ scattering amplitude, and therefore in principle the LO full off-shell $nd$ scattering amplitude must be calculated.  This calculation has not been performed in \EFT for nuclear systems, but an analogous calculation of full off-shell scattering amplitudes using the $K$-matrix below the two-body breakup energy \cite{} has been performed in cold atom systems \cite{Ji:2012nj}.  Above the two-body breakup energy the $K$-matrix approach will be complicated due to moving logarithmic singularities.  These singularities can be dealt with by the Hetherington-Schick method.  However, the position of the singularities in the Hetherington-Schick method for the full off-shell scattering amplitude have not been considered.  The perturbative approach of Vanasse \cite{Vanasse:2013sda,Vanasse:2015fph} allows the Hetherington-Schick method to be used to calculate diagrams with full off-shell scattering amplitudes.

\begin{figure}[th]
\centerline{\includegraphics[width=100mm]{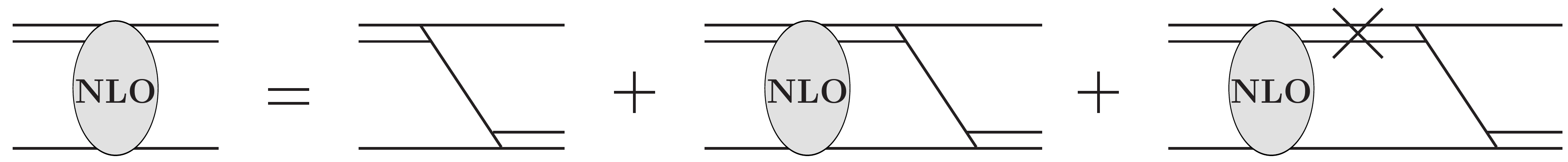}}
\caption{\label{fig:PartialResummation}Integral equation for NLO $nd$ scattering amplitude in partial resummation technique.}
\end{figure}

In order to circumvent the need to calculate the full off-shell scattering amplitude, the partial resummation technique was created\cite{Bedaque:2002yg}.  This technique is no more numerically expensive than calculating the half off-shell scattering amplitude and gives the perturbative corrections up to the order one is working.  However, one issue of the partial resummation technique is that it introduces a subset of higher order diagrams and is thus not strictly perturbative.  Also it is found for the quartet $S$-wave phase shift that above the deuteron breakup threshold the imaginary part of the NLO phase shift is negative, which is unphysical.  The NLO $nd$ scattering amplitude in the partial resummation technique is given in Fig.~\ref{fig:PartialResummation},  where a term with a single effective range insertion is added to the kernel of the integral equation.  This is equivalent to replacing the LO dibaryon propagator in the LO integral equation with the NLO dibaryon propagator.  Upon iteration of the integral equation represented in Fig.~\ref{fig:PartialResummation} the LO $nd$ scattering amplitude is obtained, and the NLO correction in Fig.~\ref{fig:NLOCorrectionDiagram}, but also the second diagram in Fig.~\ref{fig:NNLOCorrectionDiagram} and an infinite set of diagrams with single effective range insertions between $nd$ scattering amplitudes. 
This technique has been used to calculate phase shifts in $nd$ as well as $pd$ scattering\cite{Bedaque:2002yg,Griesshammer:2004pe,Konig:2011yq}.  A modification of this technique resums all effective range corrections into the dibaryon propagator.  This is certainly not perturbative with respect to the effective range insertion, but rather treats it non-perturbatively\cite{Gabbiani:2001yh,Griesshammer:2004pe}.  This method introduces a dibaryon propagator with a denominator quadratic in momentum.  The quadratic creates two poles, one the physical deuteron pole, and the other a spurious bound state pole.  Although this spurious pole is outside the range of validity of \EFT it introduces numerical difficulties in the Hetherington-Schick method and can still noticeably influence physics in the range of validity of \EFT\cite{Griesshammer:2004pe}.

\begin{figure}[th]
\centerline{\includegraphics[width=100mm]{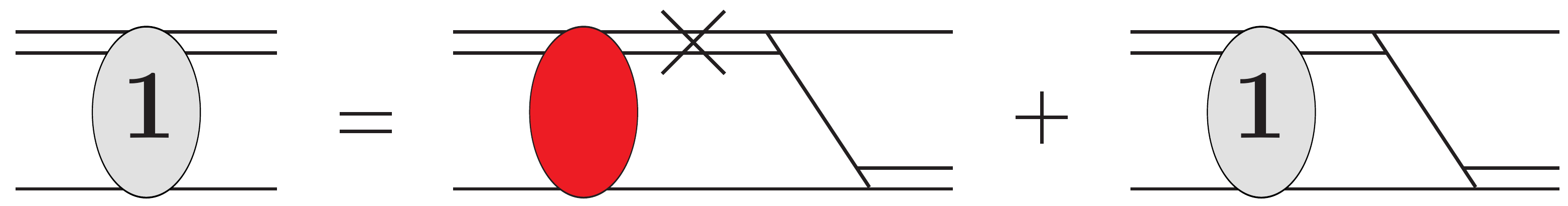}}
\caption{\label{fig:NLOPerturbativeTechniqueOld} Integral equation for NLO correction to $nd$ scattering amplitude in quartet channel.  The oval with a ``1" represents the NLO correction $t_{\mathrm{NLO}}$.}
\end{figure}

A technique to calculate the $nd$ scattering amplitude strictly perturbatively that is no more numerically expensive than calculating the half off-shell scattering amplitude was given in Ref.~\refcite{Vanasse:2013sda}.  The NLO correction to the $nd$ scattering amplitude in this technique is given in Fig.~\ref{fig:NLOPerturbativeTechniqueOld}, where the oval with a ``1" represents the NLO correction.  
The effective range insertion is now moved to the inhomogeneous part of the integral equation where it is integrated with the half-off shell LO scattering amplitude.  This integral equation gives the diagram in Fig.~\ref{fig:NLOCorrectionDiagram} and only this diagram.  The kernel for this integral equation is exactly the same kernel for the LO integral equation and this is also true at higher orders in this technique.  The power of this technique comes from the fact that whatever is put in the inhomogeneous term will simply get an additional LO $nd$ scattering amplitude attached to it.  Thus this technique can also be used for diagrams with external currents.

\begin{figure}[th]
\centerline{\includegraphics[width=100mm]{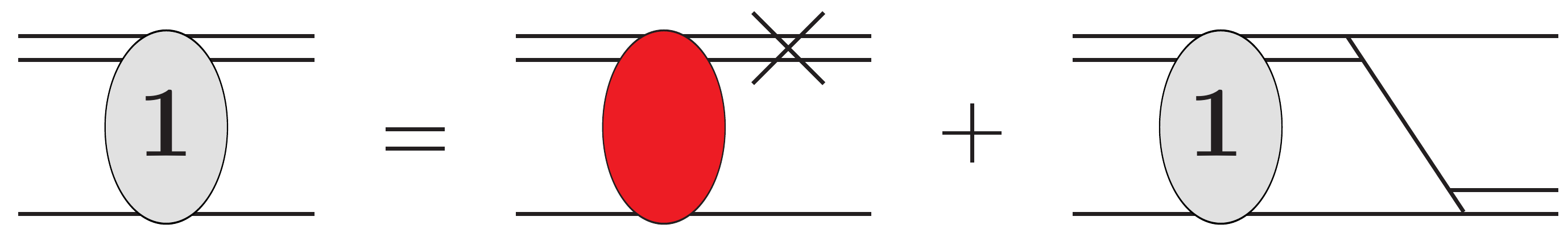}}
\caption{\label{fig:NLOPerturbativeTechniqueNew} Improved integral equation for NLO correction to $nd$ scattering amplitude in quartet channel.  The oval with a ``1" represents the NLO correction $T_{\mathrm{NLO}}$.}
\end{figure}

This perturbative technique was improved upon slightly in Ref.~\refcite{Vanasse:2015fph}, where the NLO correction to the  $nd$ scattering amplitude is given by the integral equation in Fig.~\ref{fig:NLOPerturbativeTechniqueNew}.  
The only difference between Figs.~\ref{fig:NLOPerturbativeTechniqueOld} and \ref{fig:NLOPerturbativeTechniqueNew} is a single nucleon exchange.  The lack of the single nucleon exchange means that an integration over a loop is now traded for a simple multiplication of the LO $nd$ scattering amplitude by an effective range insertion, and this introduces a slight numerical efficiency.  Upon iterating this integral equation all the diagrams in Fig.~\ref{fig:NLOPerturbativeTechniqueOld} are obtained plus the single inhomogeneous diagram in Fig.~\ref{fig:NLOPerturbativeTechniqueNew}.  When put full on-shell the inhomogeneous contribution of Fig.~\ref{fig:NLOPerturbativeTechniqueNew} gives the LO $nd$ scattering amplitude times the NLO deuteron wavefunction renormalization.  Thus in the on-shell limit the newer perturbative technique gives $T_{\mathrm{NLO}}=Z_{\mathrm{LO}}t_{\mathrm{NLO}}+Z_{\mathrm{NLO}}t_{\mathrm{LO}}$ whereas the old perturbative technique only gives $Z_{\mathrm{LO}}t_{\mathrm{NLO}}$.  Therefore, the new perturbative technique automatically gives the full NLO correction to the $nd$ scattering amplitude.  The NLO correction to the $nd$ scattering amplitude is given by
\begin{equation}
t_{1,q}^{\ell}(k,p)=t_{0,q}^{\ell}(k,p)R_{1}(p,E)+K_{0}^{\ell}(q,p,E)\otimes t_{1,q}^{\ell}(k,q),
\end{equation}
where the 0 (1,2,...) subscript means LO (NLO,N$^{2}$LO,...), and
\begin{equation}
K_{0}^{\ell}(q,p,E)=\frac{1}{\sqrt{\frac{3}{4}q^{2}-M_{N}E-i\epsilon}-\gamma_{t}}\frac{1}{qp}Q_{\ell}\left(\frac{p^{2}+q^{2}-M_{N}E-i\epsilon}{pq}\right)
\end{equation}
is the LO kernel.  The ``$\otimes$'' operator is defined by
\begin{equation}
A(q)\otimes B(q)=\frac{2}{\pi}\int_{0}^{\Lambda}dqq^{2}A(q)B(q),
\end{equation}
where $\Lambda$ is a cutoff used to regulate potential divergences and used in numerical calculations.  $R_{1}(p,E)$, the effective range insertion term is independent of the given partial wave ``$\ell$'' and is
\begin{equation}
\label{eq:R1}
R_{1}(p,E)=
\frac{Z_{t}-1}{2\gamma_{t}}\left(\gamma_{t}+\sqrt{\frac{3}{4}p^{2}-M_{N}E-i\epsilon}\right).
\end{equation}
%

\begin{figure}[th]
\centerline{\includegraphics[width=100mm]{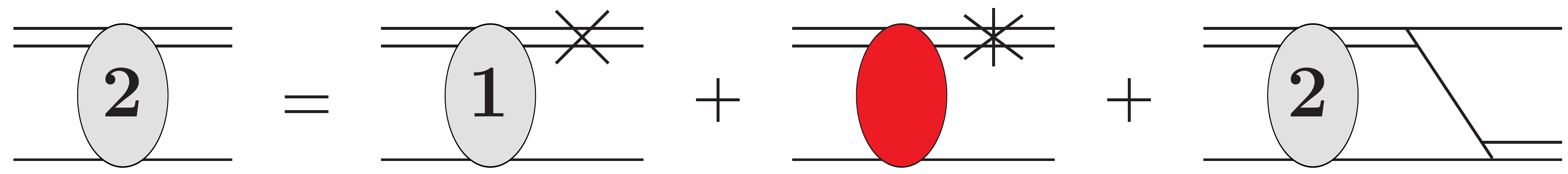}}
\caption{\label{fig:NNLOPerturbativeTechniqueNew} Integral equation for N$^{2}$LO correction to $nd$ scattering amplitude in quartet channel.  The oval with a ``2" represents the N$^{2}$LO correction.}
\end{figure}

At N$^{2}$LO the correction to the $nd$ scattering amplitude is given by the integral equation in Fig.~\ref{fig:NNLOPerturbativeTechniqueNew}.
The second inhomogeneous term contains a higher order effective range correction, $c_{0t}^{(1)}$ to ensure that the deuteron pole has the same residue.  In the ERE parametrization this second inhomogeneous term does not appear.  From Fig.~\ref{fig:NNLOPerturbativeTechniqueNew} we obtain the integral equation
\begin{equation}
t_{2,q}^{\ell}(k,p)=\left[t_{1,q}^{\ell}(k,p)-(Z_{t}-1)t_{0,q}^{\ell}(k,p)\right]R_{1}(p,E)+K_{0}^{\ell}(q,p,E)\otimes t_{2,q}^{\ell}(k,q),
\end{equation}
where $R_{1}(p,E)$ is defined in Eq. (\ref{eq:R1}).

\subsection{Doublet Channel}

$nd$ scattering in the doublet channel is entirely analogous to the quartet channel, with two extra complications.  Firstly, the doublet channel now has two coupled integral equations because the neutron and spin-singlet dibaryon can couple to give $S=1/2$, as well as the neutron and deuteron.  Using the cluster configuration space formalism \cite{Griesshammer:2004pe} the coupled equations for the doublet channel can be cast in a form similar to the quartet channel.  Secondly, the doublet $S$-wave channel contains three-body forces at all orders.  At LO the doublet $S$-wave requires a three-body contact force with no derivatives that receives corrections at each order\cite{Bedaque:1999ve,Bedaque:2002yg}.  A new energy dependent three-body force first occurs at N$^{2}$LO\cite{Bedaque:2002yg}.

\begin{figure}[th]
\centerline{\includegraphics[width=100mm]{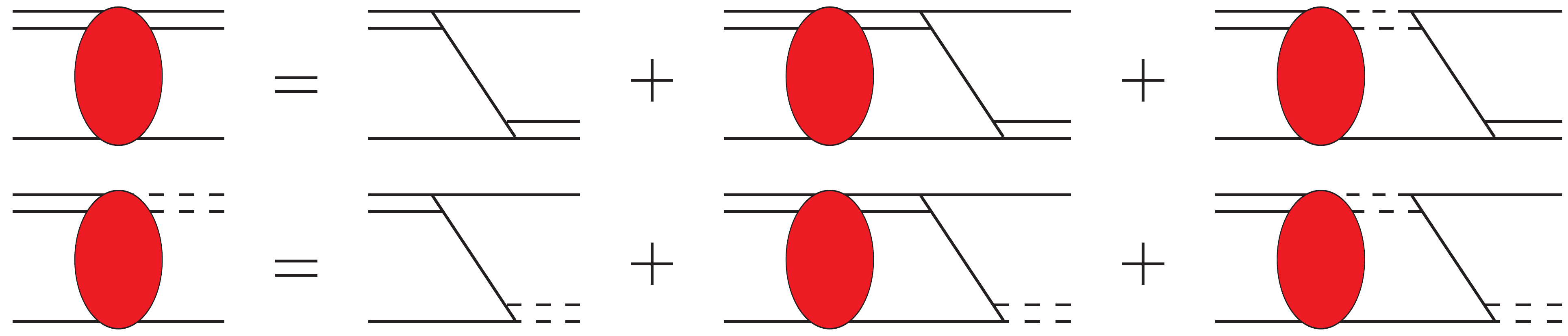}}
\caption{\label{fig:DoubletLO} Coupled integral equations for LO $nd$ scattering amplitude in doublet channel.}
\end{figure}

\begin{figure}[th]
\centerline{\includegraphics[width=100mm]{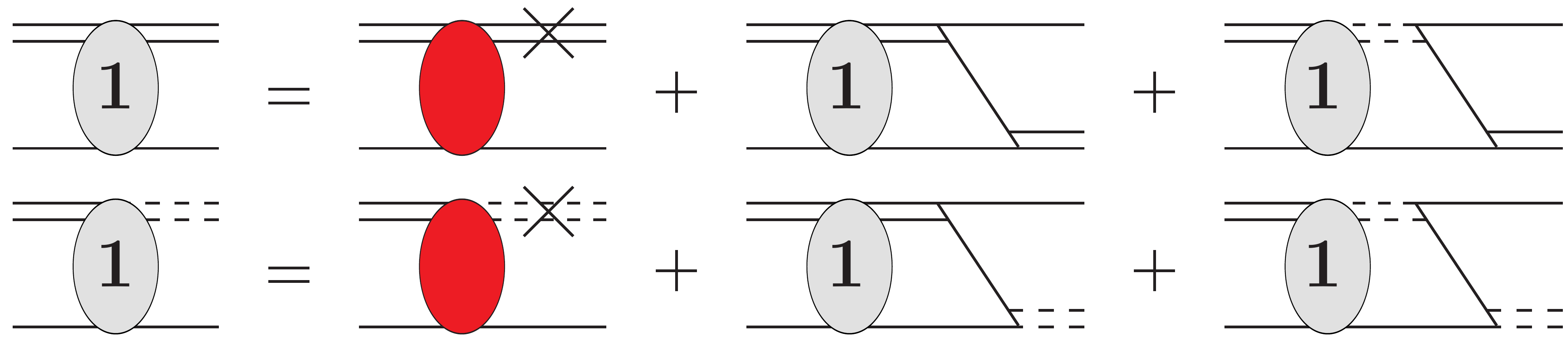}}
\caption{\label{fig:DoubletNLO} Coupled integral equations for NLO correction to $nd$ scattering amplitude in doublet channel.}
\end{figure}

\begin{figure}[th]
\centerline{\includegraphics[width=125mm]{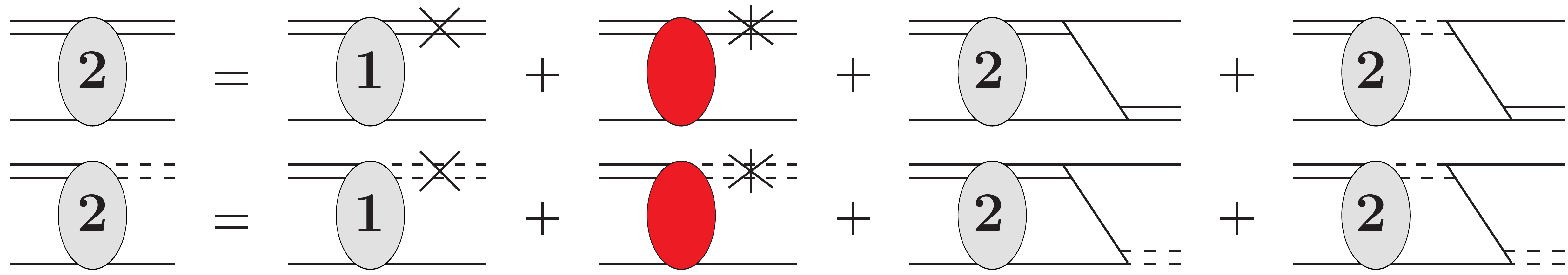}}
\caption{\label{fig:DoubletNNLO} Coupled integral equations for N$^{2}$LO correction to $nd$ scattering amplitude in doublet channel.}
\end{figure}

The LO $nd$ scattering amplitude in the doublet channel, with the exception of the $S$-wave, is given by the set of coupled integral equations in Fig.~\ref{fig:DoubletLO}, where the double-dashed line represents a spin-singlet dibaryon propagator.  
Using the cluster-configuration space \cite{Griesshammer:2004pe} formalism the integral equations can be written as a matrix equation yielding
\begin{equation}
\label{eq:DoubletLO}
\mathbf{t}_{0,d}^{\ell}(k,p)=\mathbf{B}_{0}^{\ell}(k,p,E)+\mathbf{K}_{0}^{\ell}(q,p,E)\otimes \mathbf{t}_{0,d}^{\ell}(k,q),
\end{equation}
where $\mathbf{t}_{0,d}^{\ell}(k,p)$ and $\mathbf{B}_{0}^{\ell}(k,p,E)$ are vectors in cluster configuration space defined by
\begin{equation}
\mathbf{t}_{n,d}(k,p)=
\left(\begin{array}{c}
t_{n,Nt\to Nt}^{\ell}(k,p)\\
t_{n,Nt\to Ns}^{\ell}(k,p)
\end{array}\right), 
\,\,\mathbf{B}_{0}^{\ell}(k,p,E)=
\left(\begin{array}{c}
\frac{2\pi}{pk}Q_{\ell}\left(\frac{p^{2}+k^{2}-M_{N}E-i\epsilon}{pk}\right) \\
-\frac{6\pi}{pk}Q_{\ell}\left(\frac{p^{2}+k^{2}-M_{N}E-i\epsilon}{pk}\right) 
\end{array}\right).
\end{equation}
Here $t_{n,Nt\to Nt}^{\ell}(k,p)$ ($t_{n,Nt\to Ns}^{\ell}(k,p)$) is the $n$'th order amplitude for $nd$ scattering ($nd$ going to a nucleon and spin-singlet dibaryon).  The LO kernel is a matrix in cluster configuration space defined by
\begin{align}
&\mathbf{K}_{0}^{\ell}(q,p,E)=\\\nonumber
&\hspace{.5cm}\frac{1}{2qp}Q_{\ell}\left(\frac{p^{2}+q^{2}-M_{N}E-i\epsilon}{pq}\right)
\left(\begin{array}{cc}
\frac{1}{\sqrt{\frac{3}{4}q^{2}-M_{N}E-i\epsilon}-\gamma_{t}} & \frac{-3}{\sqrt{\frac{3}{4}q^{2}-M_{N}E-i\epsilon}-\gamma_{s}} \\
\frac{-3}{\sqrt{\frac{3}{4}q^{2}-M_{N}E-i\epsilon}-\gamma_{t}} & \frac{1}{\sqrt{\frac{3}{4}q^{2}-M_{N}E-i\epsilon}-\gamma_{s}}
\end{array}\right).
\end{align}
The NLO and N$^{2}$LO correction to $nd$ scattering in the doublet channel are given by the coupled integral equations in Fig.~\ref{fig:DoubletNLO} and Fig.~\ref{fig:DoubletNNLO} respectively.
In  analogy with the quartet channel the NLO scattering amplitude in cluster configuration space is given by
\begin{equation}
\mathbf{t}_{1,d}^{\ell}(k,p)=\mathbf{t}_{0,d}^{\ell}(k,p)\circ\mathbf{R}_{1}(p,E)+\mathbf{K}_{0}^{\ell}(q,p,E)\otimes \mathbf{t}_{1,d}^{\ell}(k,q),
\end{equation}
and the N$^{2}$LO amplitude by
\begin{equation}
\mathbf{t}_{2,d}^{\ell}(k,p)=\left[\mathbf{t}_{1,d}^{\ell}(k,p)-\mathbf{c}_{1}\circ\mathbf{t}_{0,d}^{\ell}(k,p)\right]\circ\mathbf{R}_{1}(p,E)+\mathbf{K}_{0}^{\ell}(q,p,E)\otimes \mathbf{t}_{2,d}^{\ell}(k,q).
\end{equation}
The cluster configuration space vectors $\mathbf{c}_{1}$ and $\mathbf{R}_{1}(p,E)$ are given by
\begin{equation}
\mathbf{c}_{1}=
\left(\begin{array}{c}
Z_{t}-1\\
Z_{s}-1
\end{array}\right),
\end{equation}
and
\begin{equation}
\mathbf{R}_{1}(p,E)=
\left(\begin{array}{c}
\frac{Z_{t}-1}{2\gamma_{t}}\left(\gamma_{t}+\sqrt{\frac{3}{4}p^{2}-M_{N}E-i\epsilon}\right)\\
\frac{Z_{s}-1}{2\gamma_{s}}\left(\gamma_{s}+\sqrt{\frac{3}{4}p^{2}-M_{N}E-i\epsilon}\right)
\end{array}\right).
\end{equation}
The symbol ``$\circ$'' represents the Schur product of two vectors, which is simply element wise matrix multiplication.

\subsubsection{Doublet $S$-wave}
\begin{figure}[th]
\centerline{\includegraphics[width=80mm]{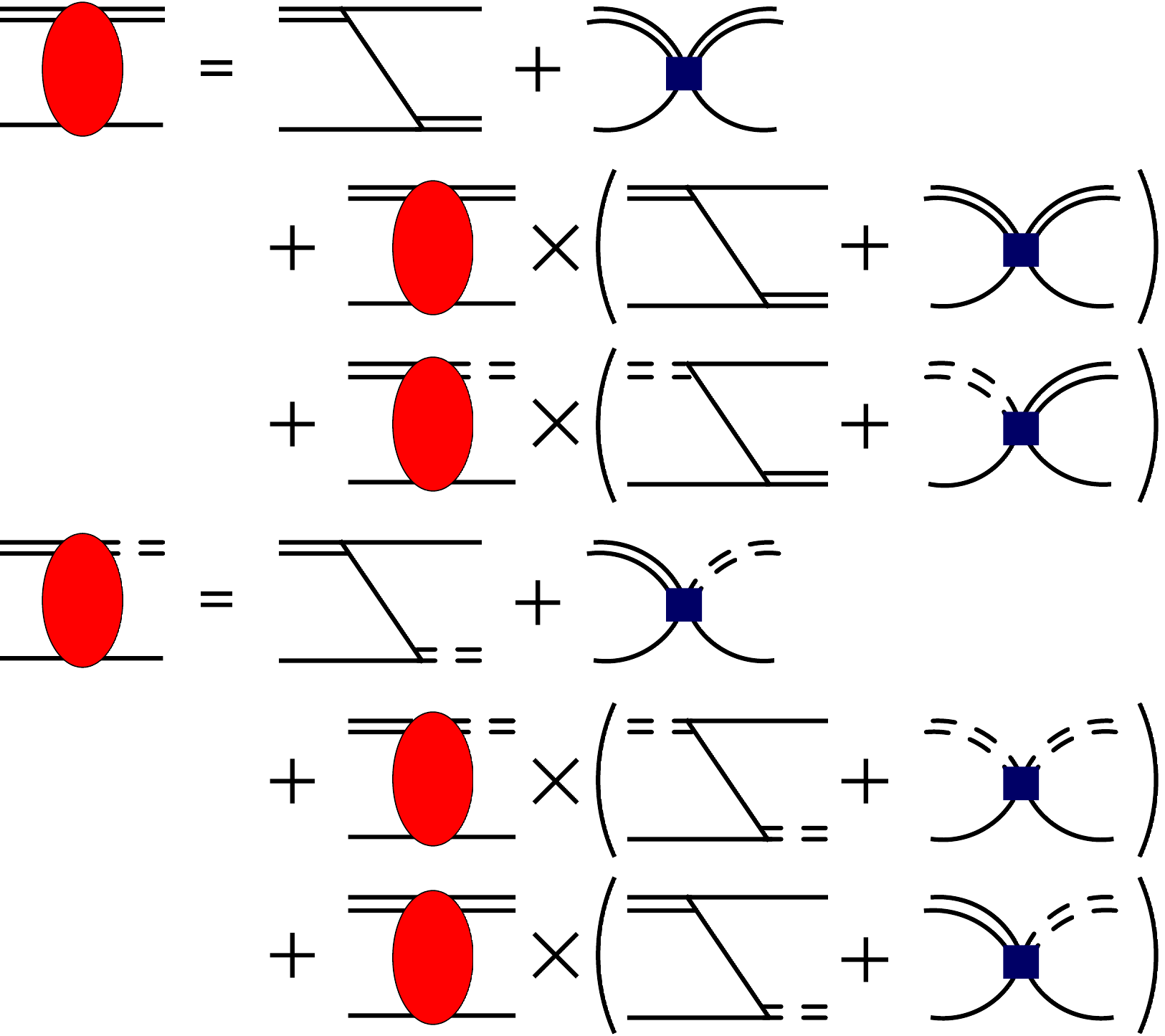}}
\caption{\label{fig:DoubletLOSwave} Coupled integral equations for LO $nd$ scattering amplitude in doublet $S$-wave channel including three-body forces.  The solid square represents the three-body force.}
\end{figure}

The LO doublet $S$-wave amplitude requires the insertion of a three-body force and is given by the set of coupled integral equations in Fig.~\ref{fig:DoubletLOSwave}.
The solid square represents the LO three-body force given by the Lagrangian
\begin{align}
&{\mathcal{L}}_{3}=\frac{M_{N}H_{0}(\Lambda)}{3\Lambda^{2}}\left[y_{t}\hat{N}^{\dagger}(\vec{t}\cdot\vectS{\sigma})^{\dagger}-y_{s}\hat{N}^{\dagger}(\vec{s}\cdot\vectS{\tau})^{\dagger}\right]
\left[y_{t}(\vec{t}\cdot\vectS{\sigma})\hat{N}-y_{s}(\vec{s}\cdot\vectS{\tau})\hat{N}\right].
\end{align}
Unlike in the quartet channel the Pauli principle does not prevent all three particles from meeting at a point and therefore the doublet $S$-wave channel is sensitive to short distance physics that is encoded in the three-body force.  In the limit where $\Lambda\to\infty$ the integral equation without a three-body force does not posses a unique solution but instead has an arbitrary phase\cite{Bedaque:1998kg,Bedaque:1998km,Bedaque:1999ve}.  For finite values of $\Lambda$ this results in a large $\Lambda$ dependence since it cannot converge to a unique solution.\footnote{Note if everything is properly renormalized the regulator dependence should be removed and a unique solution should be found as $\Lambda\to\infty$.}  The three-body force fixes the phase and provides a unique solution.  Further insight is gained by transforming to the Wigner basis, defined by  
\begin{equation}
\left(\begin{array}{c}
t^{(-)}(k,p)\\
t^{(+)}(k,p)
\end{array}\right)
=
\left(\begin{array}{c}
t_{n,Nt\to Nt}^{\ell=0}(k,p)-t_{n,Nt\to Ns}^{\ell=0}(k,p)\\
t_{n,Nt\to Nt}^{\ell=0}(k,p)+t_{n,Nt\to Ns}^{\ell=0}(k,p)
\end{array}\right)
\end{equation}
In the Wigner limit ($\gamma_{t}=\gamma_{s}$) the equations for $t^{(+)}(k,p)$ and $t^{(-)}(k,p)$ decouple.  The integral equation for $t^{(-)}(k,p)$ is equivalent to a three-boson problem and requires a three-body force for renormalization,  while the integral equation for $t^{(+)}(k,p)$ is the same as the quartet channel and requires no three-body force.  Calculating the asymptotic form of $t^{(-)}(k,p)$, predictions for the running of the three-body force have been made and match well to numerical calculations\cite{Bedaque:1999ve}.  Going to the Wigner-basis also shows that the LO three-body force in the doublet $S$-wave channel is Wigner symmetric\cite{Mehen:1999qs,PhysRev.51.106}.  In fact it can be shown that the only LO three-body force with no derivatives is a Wigner-symmetric three-body force in the doublet $S$-wave channel\cite{Bedaque:1999ve,Mehen:1999qs}.  At higher orders this three-body force will receive corrections and at N$^{2}$LO there is a new energy dependent three-body force\cite{Bedaque:2002yg}.  Typically the LO three-body force and its higher order corrections are fit to reproduce the doublet $S$-wave $nd$-scattering length, and the energy dependent N$^{2}$LO three-body force is fit to the triton binding energy. In order to deal with these three-body forces a new but analytically equivalent approach is used by introducing a triton auxiliary field\cite{Vanasse:2015fph}.

\section{\label{sec:boundstates}Bound States}

There has been less progress in studying the bound state regime than in the scattering regime.  Calculations of the triton binding energy have been performed at LO\cite{Bedaque:1999ve}, and the triton charge radius has also been calculated at LO \cite{Platter:2005sj,Vanasse:2015fph}.  The $nd$ capture process in both the parity-conserving (PC) and PV sector has been calculated\cite{Arani:2011if,Arani:2014qsa}.  Also the $\jjvH$~-~$\jjvHe$ binding energy difference has been calculated in \EFT with perturbative\cite{Konig:2011yq,Konig:2014ufa} and non-perturbative\cite{Ando:2010wq} treatments of Coulomb forces.  In halo EFT the introduction of an effective trimer auxiliary field by Hagen et al. was used to calculate the charge form factor of halo nuclei\cite{Hagen:2013xga}.  Building upon this work Vanasse showed a simple procedure by which perturbative corrections could be added to bound state calculations\cite{Vanasse:2015fph}.  Using this he calculated the triton charge radius to NLO.  The essential improvement on the work of Hagen et al. is the realization that certain quantities can be calculated by direct numerical integration rather than taking a numerical limiting procedure about the bound state pole.

\begin{figure}[th]
\centerline{\includegraphics[width=80mm]{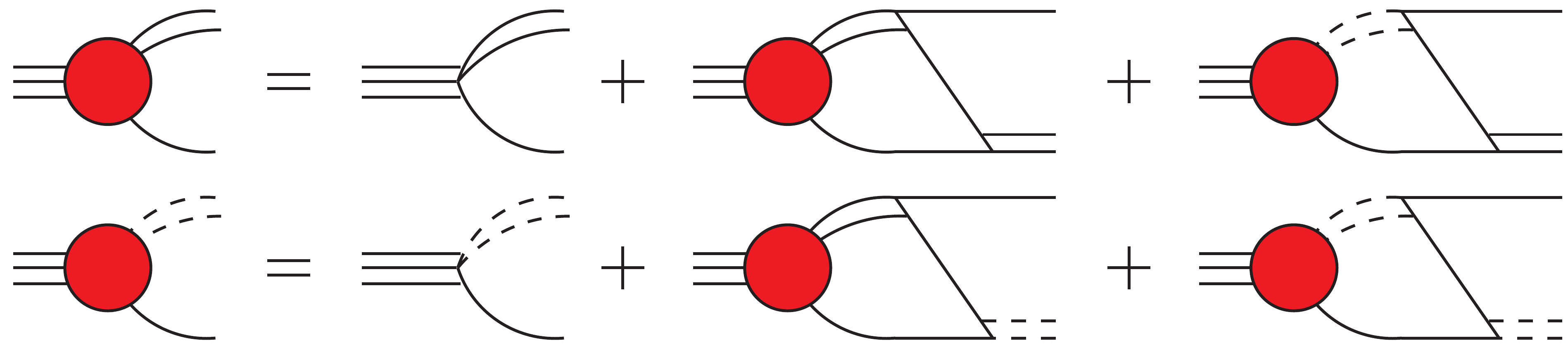}}
\caption{\label{fig:GirrLO} Coupled integral equations for LO triton vertex function.}
\end{figure}

\begin{figure}[th]
\centerline{\includegraphics[width=80mm]{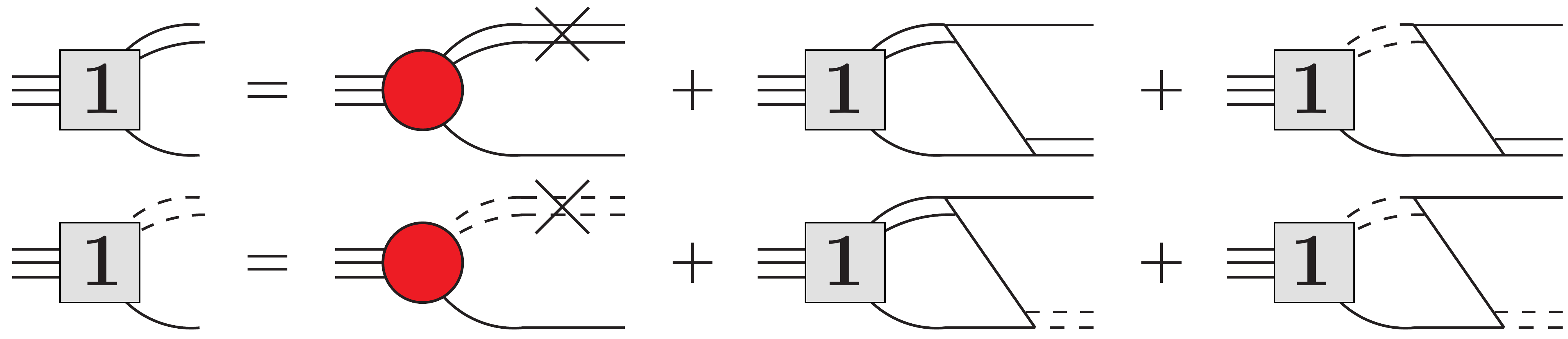}}
\caption{\label{fig:GirrNLO} Coupled integral equations for NLO correction to the triton vertex function.}
\end{figure}

\begin{figure}[th]
\centerline{\includegraphics[width=95mm]{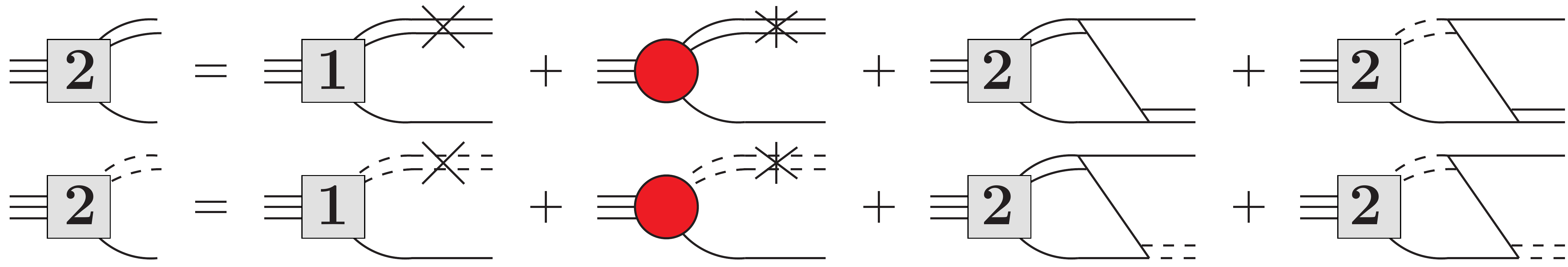}}
\caption{\label{fig:GirrNNLO} Coupled integral equations for N$^{2}$LO correction to the triton vertex function.}
\end{figure}

\begin{figure}[th]
\centerline{\includegraphics[width=80mm]{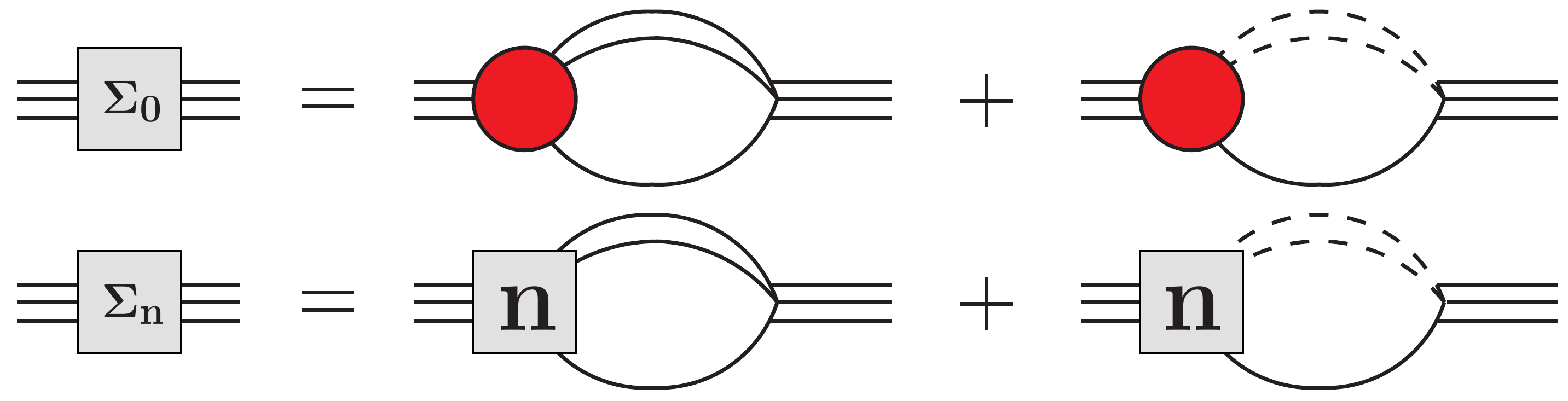}}
\caption{\label{fig:Sigma0} Diagrams for function $\Sigma_{n}(E)$}
\end{figure}

\begin{figure}[th]
\centerline{\includegraphics[width=80mm]{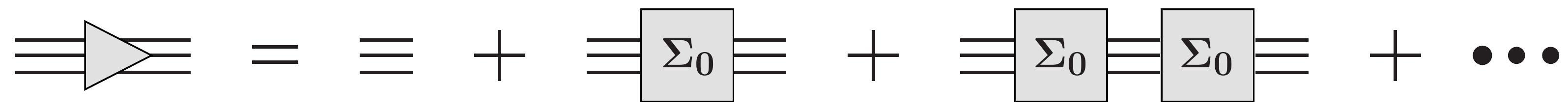}}
\caption{\label{fig:TritonProp} LO dressed triton propagator}
\end{figure}

Introducing a triton auxiliary field $\hat{\psi}$ we find the three-body Lagrangian
\begin{equation}
\label{eq:tritonLag}
\mathcal{L}_{3}=\hat{\psi}^{\dagger}\left(\Omega-h_{2}(\Lambda)\left(i\partial_{0}+\frac{\vect{\nabla}^{2}}{2M_{N}}+\frac{\gamma_{t}^{2}}{M_{N}}\right)\right)\hat{\psi}+\sum_{n=0}^{2}\omega_{0}^{(n)}\hat{\psi}^{\dagger}\left(\sigma_{i}\hat{N}\hat{t}_{i}-\tau_{a}\hat{N}\hat{s}_{a}\right)+\mathrm{H.c},
\end{equation}
where $\Omega$ is the bare triton propagator, $h_{2}(\Lambda)$ in front of the triton kinetic term is related to the N$^{2}$LO energy dependent three-body force, and the last term contains interactions up to N$^{2}$LO between the triton, dibaryon, and nucleon fields. Note that these interaction terms are Wigner-symmetric.  The LO triton vertex function is given by the coupled integral equations in Fig.~\ref{fig:GirrLO}, where the triple lines are triton propagators.  In cluster configuration space these integral equations are given by
\begin{equation}
\boldsymbol{\mathcal{G}}_{0}(E,p)=\widetilde{\mathbf{B}}_{0}+\mathbf{K}^{\ell=0}_{0}(q,p,E)\otimes\boldsymbol{\mathcal{G}}_{0}(E,q),
\end{equation}
where
\begin{equation}
\boldsymbol{\mathcal{G}}_{n}(E,p)=
\left(\begin{array}{c}
\mathcal{G}_{n,\psi\to Nt}(E,p)\\
\mathcal{G}_{n,\psi\to Ns}(E,p)
\end{array}\right),
\widetilde{\mathbf{B}}_{0}=
\sqrt{3}\omega_{0}^{(0)}\left(\!\begin{array}{r}
1\\
-1
\end{array}\right).
\end{equation}
The only difference between the integral equation for the LO triton vertex function and Eq. (\ref{eq:DoubletLO}) with $\ell=0$ is in the inhomogeneous term.  NLO  and N$^{2}$LO corrections to the triton vertex function are given by the integral equations in Fig.~\ref{fig:GirrNLO} and Fig.~\ref{fig:GirrNNLO} respectively.
The NLO correction to the triton vertex function is given by
\begin{equation}
\boldsymbol{\mathcal{G}}_{1}(E,p)=\boldsymbol{\mathcal{G}}_{0}(E,p)\circ\mathbf{R}_{1}(p,E)+\mathbf{K}^{\ell=0}_{0}(q,p,E)\otimes\boldsymbol{\mathcal{G}}_{1}(E,q),
\end{equation}
and the N$^{2}$LO correction by
\begin{equation}
\boldsymbol{\mathcal{G}}_{2}(E,p)=\Big[\boldsymbol{\mathcal{G}}_{1}(E,p)-\mathbf{c}_{1}\circ\boldsymbol{\mathcal{G}}_{0}(E,p)\Big]\circ\mathbf{R}_{1}(p,E)+\mathbf{K}^{\ell=0}_{0}(q,p,E)\otimes\boldsymbol{\mathcal{G}}_{2}(E,q).
\end{equation}
Again these equations are entirely analogous to those for $nd$ scattering with the only difference being the LO inhomogeneous term.  From the LO triton vertex function and its perturbative corrections
\begin{equation}
\Sigma_{n}(E)=\frac{1}{2\pi^{2}}\int_{0}^{\Lambda}dq q^{2}
\left(\begin{array}{c}
\frac{1}{\sqrt{\frac{3}{4}q^{2}-M_{N}E-i\epsilon}-\gamma_{t}}\\
\frac{1}{\sqrt{\frac{3}{4}q^{2}-M_{N}E-i\epsilon}-\gamma_{s}}
\end{array}\right)\cdot
\left(\begin{array}{c}
\mathcal{G}_{n,\psi\to Nt}(E,p)\\
\mathcal{G}_{n,\psi\to Ns}(E,p)
\end{array}\right),
\end{equation}
given in Fig.~\ref{fig:Sigma0}.
Using $\Sigma_{0}(E)$ we define the LO triton propagator by summing the diagrams in Fig.~\ref{fig:TritonProp},
which yields the LO dressed triton propagator given by
\begin{equation}
i\Delta_{3}(E)=\frac{i}{\Omega}\frac{1}{1-H_{\mathrm{LO}}\Sigma_{0}(E)}.
\end{equation}
The triton pole is given at the energy ($E=B$) for which
\begin{equation}
\Sigma_{0}(B)=\frac{1}{H_{\mathrm{LO}}}.
\end{equation}
In this way the LO three-body force, $H_{\mathrm{LO}}=-3(\omega_{0}^{(0)})^{2}/(4\pi\Omega)$, can be fit to the triton binding energy.  Taking the residue about the triton pole gives the triton wavefunction renormalization,
\begin{equation}
Z_{\psi}=-\frac{1}{\Omega}\frac{1}{H_{\mathrm{LO}}\Sigma_{0}'(B)}.
\end{equation}
Combining the LO triton vertex function with the triton wavefunction renormalization gives the properly renormalized triton vertex function.  The properly renormalized triton vertex function is equivalent to solving the homogeneous equation for doublet $S$-wave scattering with a three-body force and properly normalizing it\cite{Konig:2011yq}.

\begin{figure}[th]
\centerline{\includegraphics[width=80mm]{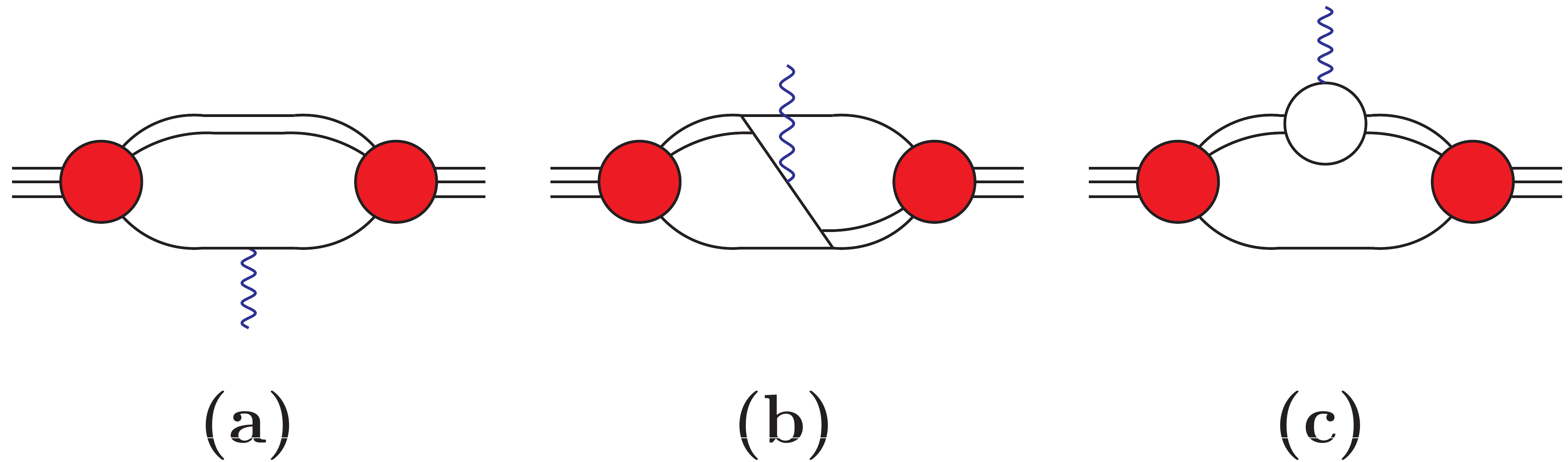}}
\caption{\label{fig:FormFactor} Diagrams for LO charge form factor of triton. The wavy lines are minimally coupled $A_{0}$ photons.}
\end{figure}

\begin{figure}[th]
\centerline{\includegraphics[width=80mm]{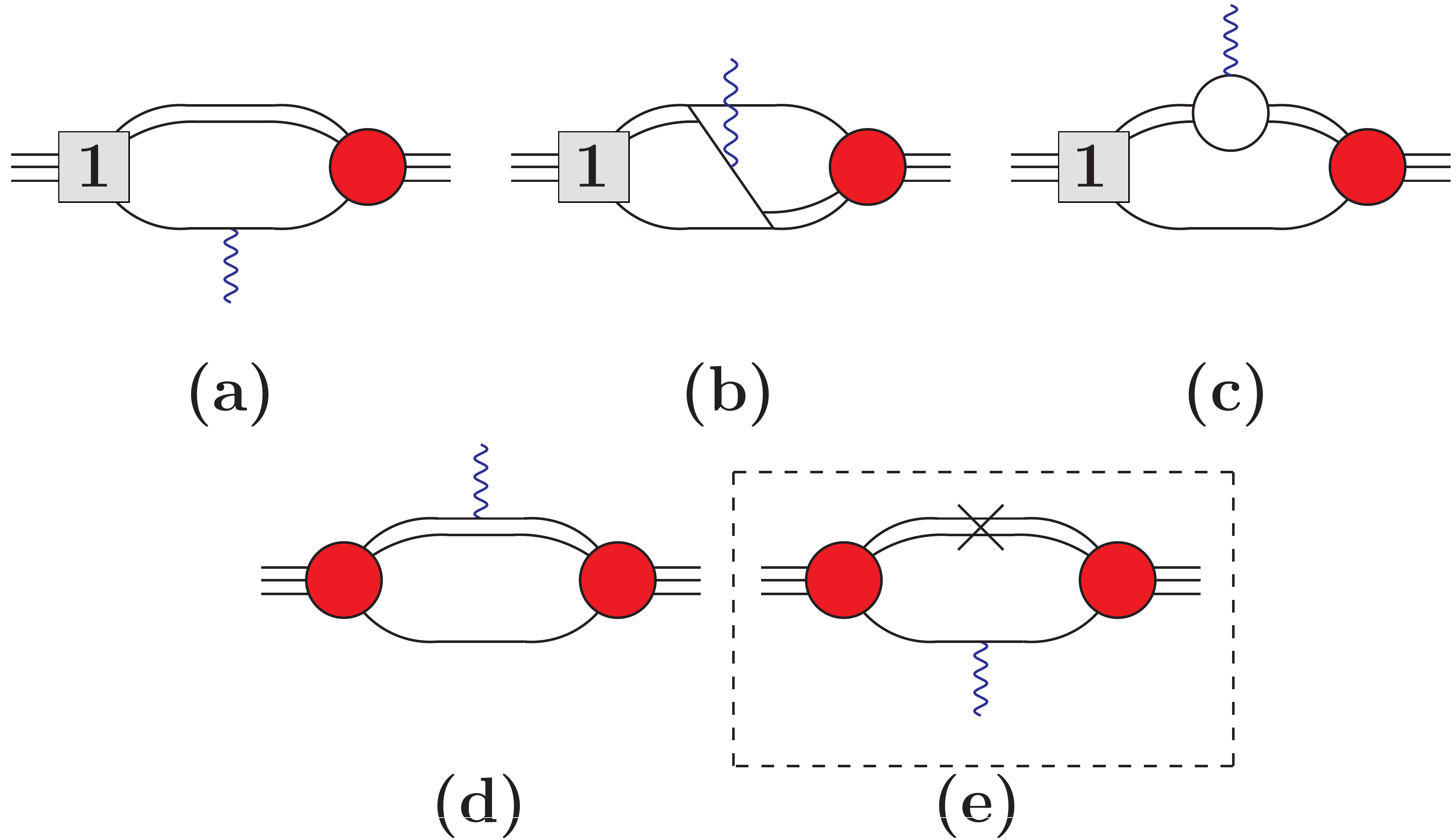}}
\caption{\label{fig:FormFactorNLO} Diagrams for the NLO correction to the charge form factor of the triton. Diagrams related by time reversal symmetry are not shown, and diagram (e) is subtracted from the other diagrams to avoid double counting.}
\end{figure}

\begin{figure}[th]
\centerline{\includegraphics[width=100mm]{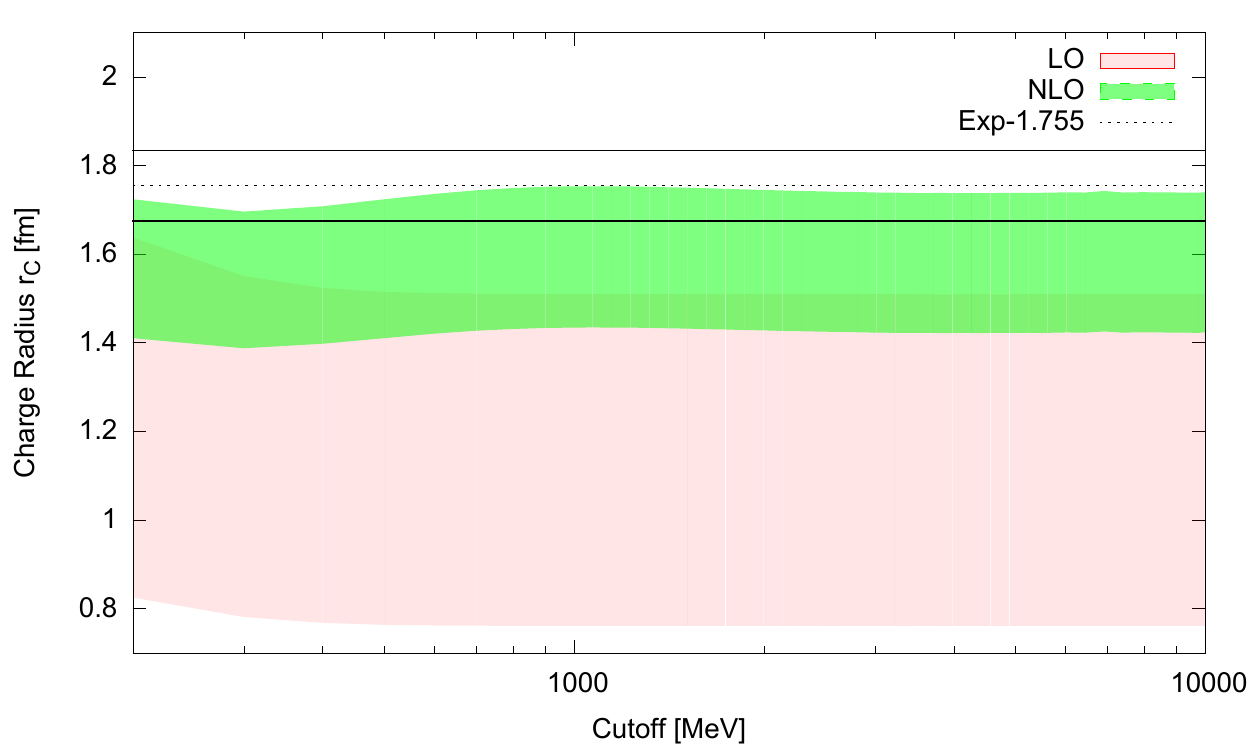}}
\caption{\label{fig:ChargeRadius}Cutoff dependence of triton charge radius in \EFT~\cite{Vanasse:2015fph}.  The pink band is a 30\% error estimate for the LO triton charge radius of $1.13$~fm and the green band a 10\% error estimate for the NLO triton charge radius of $1.59$~fm.  The dotted line is the experimental value $1.755\pm.086$~fm \cite{Amroun:1994qj} and the black lines its error.}
\end{figure}

With the triton vertex function, bound state properties of $\jjvH$ can be calculated.  For example the LO triton charge form factor is given by the diagrams in Fig.~\ref{fig:FormFactor} where the wavy lines are minimally coupled $A_{0}$ photons
and the NLO correction to the triton charge form factor is given by the diagrams in Fig.~\ref{fig:FormFactorNLO}.
Diagram (e) in the dashed box is subtracted from the other diagrams to avoid double counting.  In calculating these diagrams the triton vertex function is not in the $nd$ c.m.~frame.  The LO (NLO correction to the) triton vertex function in a boosted frame can be related to the LO (NLO correction to the) triton vertex function in the c.m.~frame via an integral equation similar to that for the LO (NLO correction to the) c.m.~triton vertex function\cite{Hagen:2013xga,Vanasse:2015fph}.  Further details of this calculation can be seen in Ref.~\refcite{Vanasse:2015fph}.  Extracting the triton charge radius from the LO and NLO correction to the triton charge form factor gives the results in Fig.~\ref{fig:ChargeRadius}.
The cutoff dependence of the LO and NLO triton charge radius is well behaved, converging at large cutoffs.  It converges to a LO value of 1.13~fm and a NLO value of 1.59~fm within 10\% of the experimental value of 1.755$\pm$.086~fm\cite{Amroun:1994qj}, where 10\% is the expected error of our NLO \EFT calculation.  The LO value is more than 30\% away from the experimental value, which is greater than the naive LO error estimate in \EFT.  A LO \EFT calculation of the triton charge radius has also been performed using a wavefunction approach, for which they find a larger value of 2.1$\pm$.6~fm\cite{Platter:2005sj}.  In addition to the triton charge form factor many other triton properties can now be calculated to higher orders as well as processes involving external currents such as $nd\to t\gamma$.

\begin{figure}[th]
\centerline{\includegraphics[width=60mm]{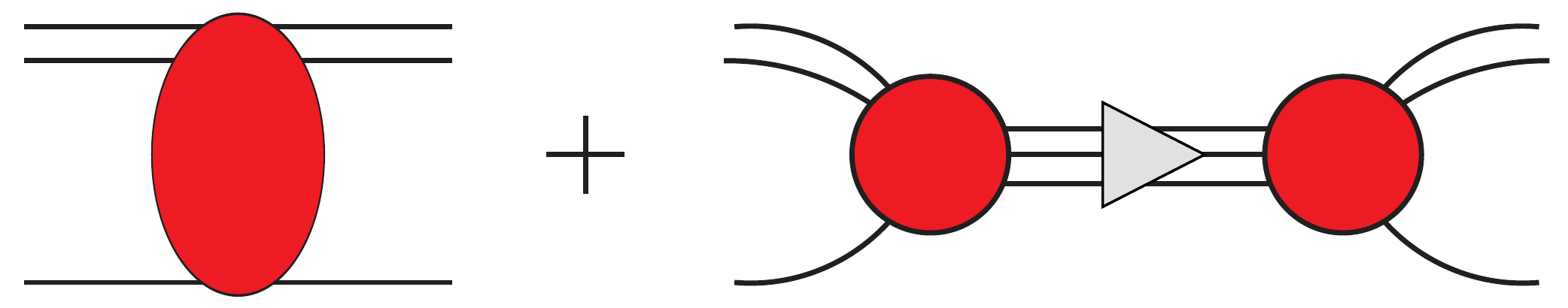}}
\caption{\label{fig:DoubletSLO} LO $nd$ scattering amplitude in doublet $S$-wave channel.}
\end{figure}

The LO triton vertex function also offers a novel way to calculate doublet $S$-wave $nd$ scattering, which at LO is given by the diagrams in Fig.~\ref{fig:DoubletSLO}.  
The first diagram is calculated from Eq. (\ref{eq:DoubletLO}) with $\ell=0$, where there is no three-body force term.  All three-body force terms are factored into the second diagram.  The sum of these two diagrams is given by the expression
\begin{equation}
T_{\mathrm{LO}}(k)=Z_{\mathrm{LO}}t_{0,Nt\to Nt}^{\ell=0}(k,k)+H_{\mathrm{LO}}\frac{1}{1-H_{\mathrm{LO}}\Sigma_{0}(E)}\pi Z_{\mathrm{LO}}(G_{0,\psi\to Nt}(E,k))^{2}.
\end{equation}
In this expression the LO three-body force is factored out of all numerically determined expressions.  Therefore, the LO three-body force can be calculated analytically in terms of numerically determined quantities, yielding
\begin{equation}
H_{\mathrm{LO}}=\frac{x}{1+x\Sigma_{0}(-\gamma_{t}^{2})},
\end{equation}
where
\begin{equation}
x=\frac{-\left(\frac{3\pi a_{nd}}{M_{N}}+Z_{\mathrm{LO}}t_{0,Nt\to Nt}^{\ell=0}(0,0)\right)}{\pi Z_{\mathrm{LO}}(G_{0,\psi\to Nt}(-\gamma_{t}^{2},0))^{2}}.
\end{equation}
Here the LO three-body force is fit so that the doublet $S$-wave $nd$ scattering length $a_{nd}=.65$~fm is reproduced.  

\begin{figure}[th]
\centerline{\includegraphics[width=90mm]{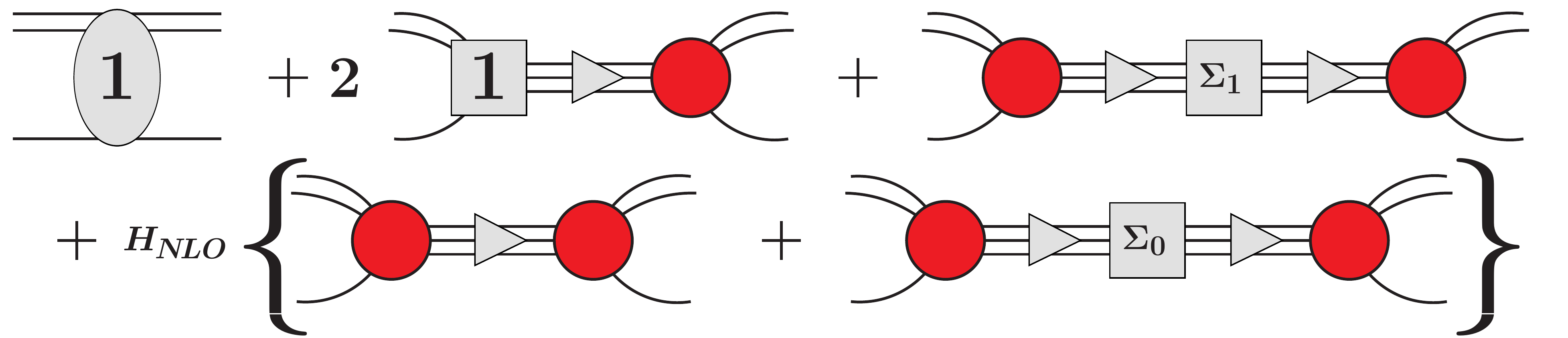}}
\caption{\label{fig:DoubletSNLO} NLO $nd$ scattering amplitude correction in doublet $S$-wave channel.}
\end{figure}

\begin{figure}[th]
\centerline{\includegraphics[width=100mm]{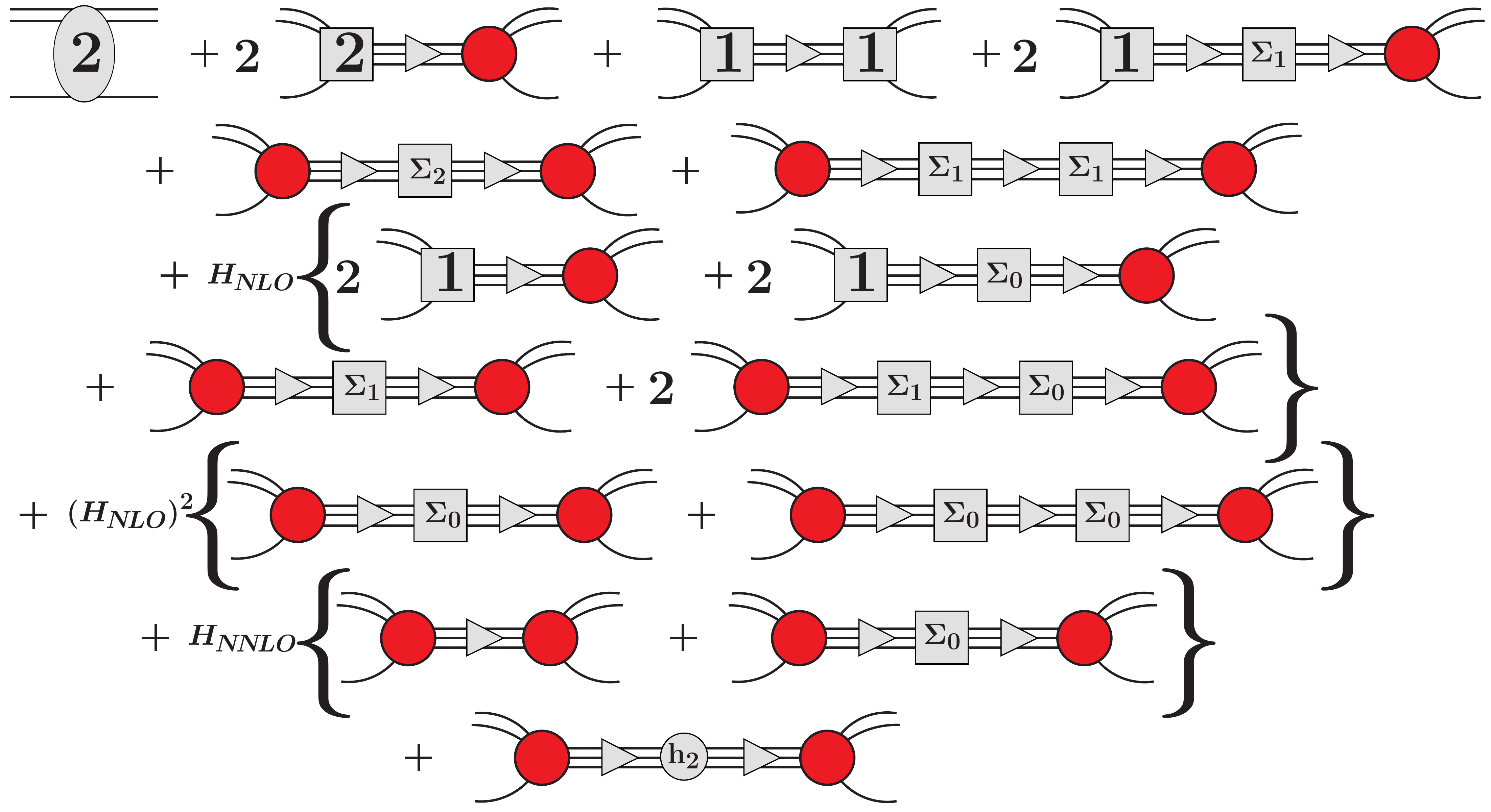}}
\caption{\label{fig:DoubletSNNLO} N$^{2}$LO $nd$ scattering amplitude correction in doublet $S$-wave channel.  The diagram with $h_{2}$ comes from the kinetic term of the triton in Eq. (\ref{eq:tritonLag}).}
\end{figure}

The NLO and N$^{2}$LO corrections to doublet $S$-wave $nd$ scattering are given by Figs.~\ref{fig:DoubletSNLO} and \ref{fig:DoubletSNNLO}, respectively.
Again for these diagrams the higher order three-body force corrections can be factored out from numerically determined quantities and analytical expressions can be written for them in terms of these numerically determined quantities.  The higher order corrections $H_{\mathrm{NLO}}$ and $H_{\mathrm{NNLO}}$ are again fit to reproduce the correct doublet $S$-wave $nd$ scattering length.  The new energy dependent three-body force $h_{2}(\Lambda)$ at N$^{2}$LO comes from the kinetic term for the triton in Eq. (\ref{eq:tritonLag}) and is fit to reproduce the triton binding energy, $E_{B}^{\jjvH}$.  The NLO correction to the triton binding energy can be shown to be
\begin{equation}
B_{1}=-\frac{H_{\mathrm{LO}}\Sigma_{1}(B_{0})+H_{\mathrm{NLO}}\Sigma_{0}(B_{0})}{H_{\mathrm{LO}}\Sigma_{0}'(B_{0})},
\end{equation}
and the N$^{2}$LO correction
\begin{align}
B_{2}=-&\frac{H_{\mathrm{LO}}\Sigma_{2}(B_{0})+H_{\mathrm{NLO}}\Sigma_{1}(B_{0})+(H_{\mathrm{NNLO}}+\frac{4}{3}\left(B_{0}+\frac{\gamma_{t}^{2}}{M_{N}}\right)H_{\mathrm{LO}}h_{2})\Sigma_{0}(B_{0})}{H_{\mathrm{LO}}\Sigma_{0}'(B_{0})}\\\nonumber
&-B_{1}\frac{H_{\mathrm{LO}}\Sigma_{1}'(B_{0})+H_{\mathrm{NLO}}\Sigma_{0}'(B_{0})}{H_{\mathrm{LO}}\Sigma_{0}'(B_{0})}-\frac{1}{2}B_{1}^{2}\frac{\Sigma_{0}''(B_{0})}{\Sigma_{0}'(B_{0})},
\end{align}
where $B_{0}$ is the LO triton binding energy when the LO three-body force is fit to the scattering length.  With these corrections $h_{2}(\Lambda)$ can be chosen such that $B_{0}+B_{1}+B_{2}=E_{B}^{\jjvH}$ and the triton binding energy is reproduced exactly at N$^{2}$LO.  Previous methods for calculating these corrections required the use of a limiting procedure\cite{Ji:2012nj}.    Calculating these quantities without the need for a limiting procedure is advantageous because it avoids the need to calculate the scattering amplitude at multiple energies and avoids errors introduced by fitting to these points calculated at multiple energies.  For a more detailed discussion of these methods see Ref.~\refcite{Vanasse:2015fph}.

\section{\label{sec:Coulomb}Coulomb forces and $pd$ scattering}

All of the techniques in $nd$ scattering can also be applied to $pd$ scattering.  However, in $pd$ scattering there is the complication of the Coulomb interaction.  The first \EFT calculations in $pd$ scattering were carried out by Rupak and Kong\cite{Rupak:2001ci}.  In their calculation they developed a new power counting in which a new scale $p$, the external momentum, was introduced.  This scale is important since in the infra-red certain diagrams scale as $1/p$ because of the Coulomb force and therefore become enhanced for small $p$.  They calculated $pd$ scattering in the quartet $S$-wave channel by treating Coulomb interactions perturbatively but resumming them to all orders in the integral equation.  In addition, they used the screening method\cite{Alt1985429,levin2013coulomb} to deal with the singularities introduced by massless photons and were only able to calculate reliably down to momenta of about 20~MeV because of numerical issues.  This work was built upon by Hammer and K{\"o}nig, which used a refined integration mesh to push to lower momenta\cite{Konig:2011yq} of about 3~MeV.  Ref.~\refcite{Konig:2011yq} also calculated the doublet $S$-wave channel phase shifts and the $\jjvHe$-$\jjvH$ binding energy difference.  However, their calculations only considered small cutoffs and therefore did not address possible issues with renormalization in the doublet $S$-wave channel for $pd$ scattering.  

\begin{figure}[!ht]
    \subfloat[\label{subfig:H1aNot}]{%
      \includegraphics[width=0.47\textwidth]{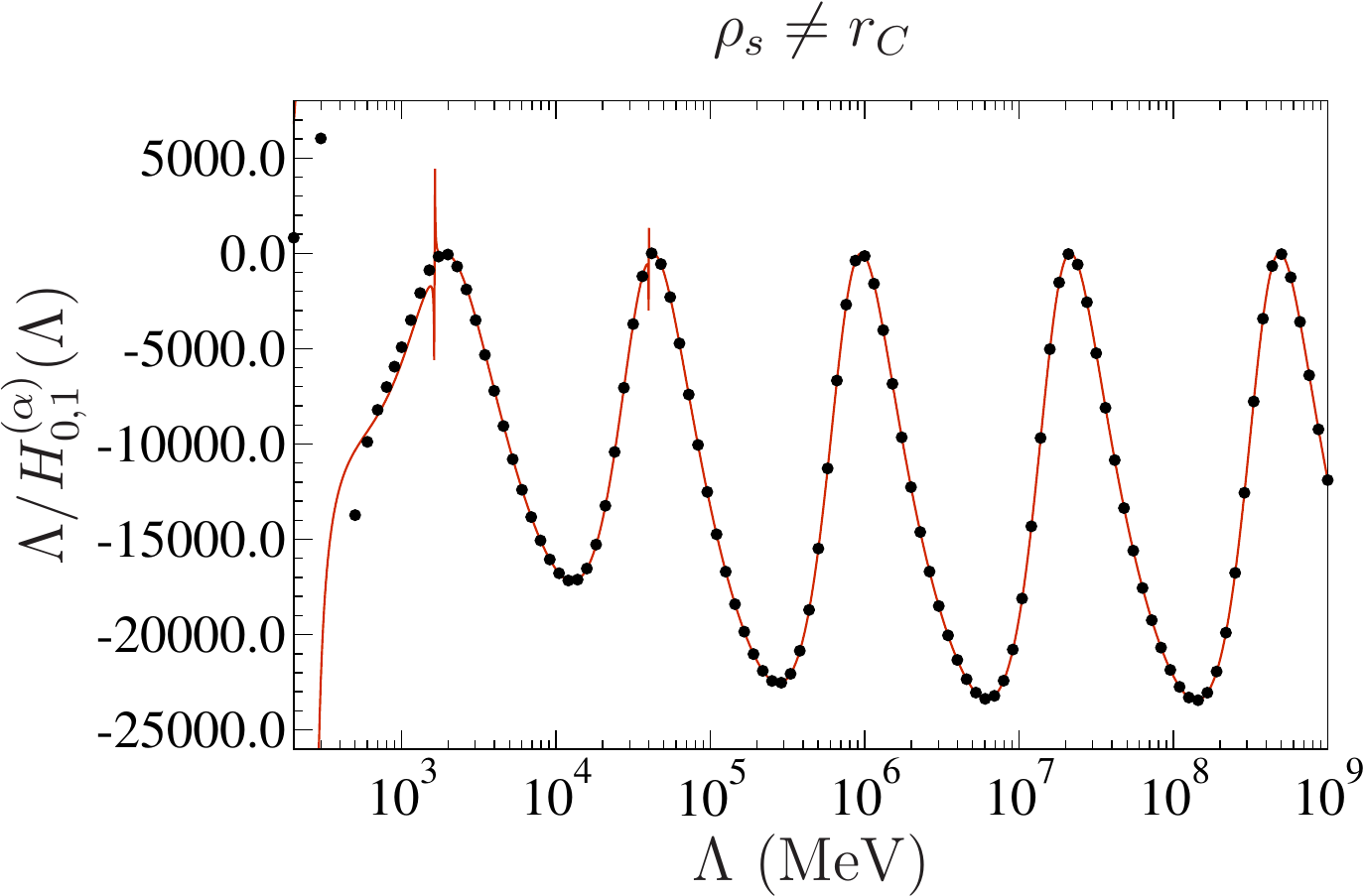}
    }
    \hfill
    \subfloat[\label{subfig:H1aIs}]{%
      \includegraphics[width=0.47\textwidth]{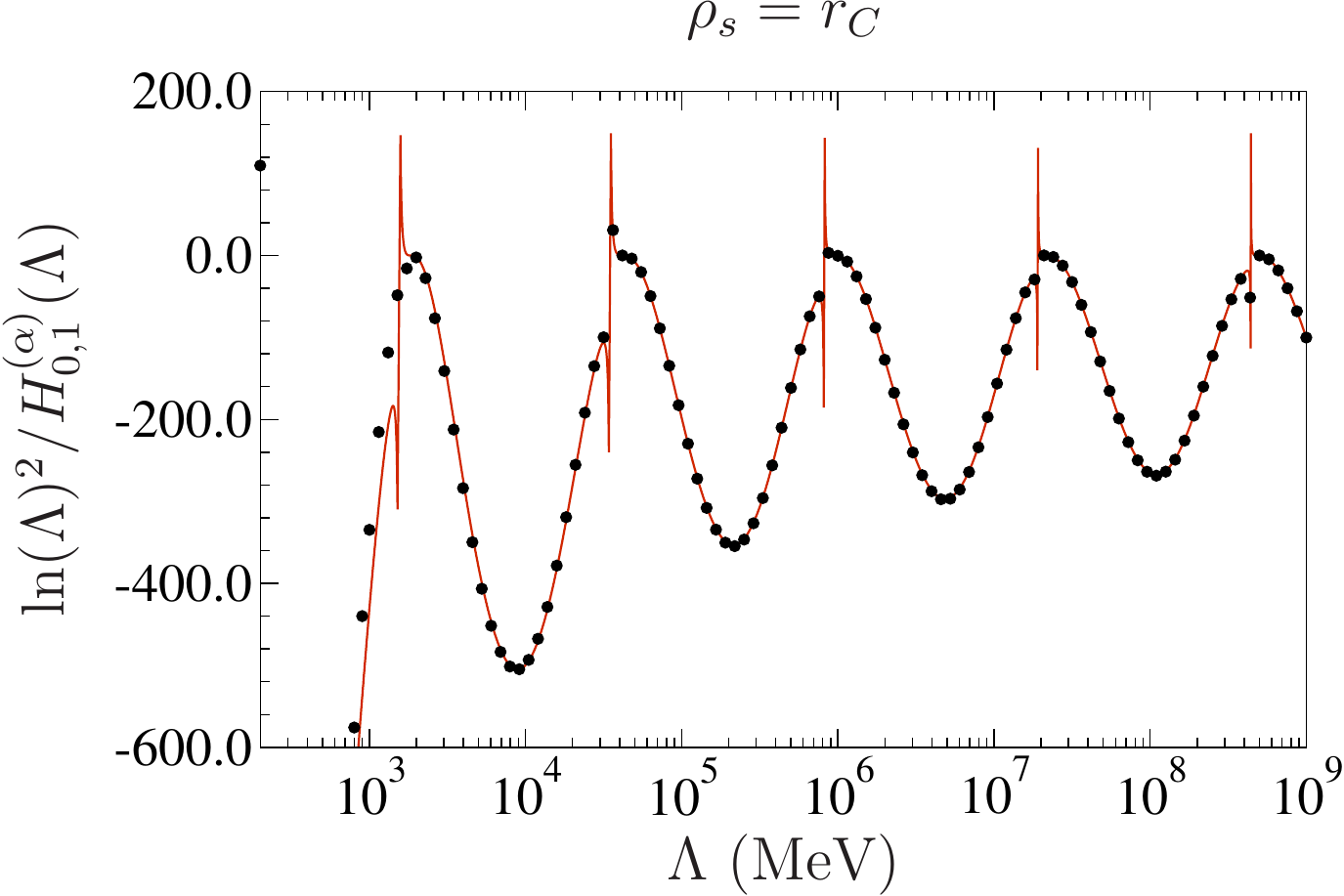}
    }
    \caption{The black dots are numerical calculations from fitting the NLO Coulomb three-body force to the $\jjvHe$ binding energy, and the solid red line analytical predictions.  Diagram (a) is for the case when $\rho_{s}\neq r_{C}$ and $\Lambda/H_{0,1}^{(\alpha)}(\Lambda)$ is plotted as a function of cutoff to divide out the linear divergence and convert poles to zeros.  Diagram (b) is for $\rho_{s}=r_{C}$ and $\ln(\Lambda)^{2}/H_{0,1}^{(\alpha)}(\Lambda)$ is plotted to divide out the dominant logarithmic behavior and convert poles to zeroes (Figures from Ref.~\citen{Vanasse:2014kxa}).}
    \label{fig:H1a}
 \end{figure}

It was later shown by Vanasse et al. that the NLO three-body force for $nd$ scattering in the doublet $S$-wave channel does not properly renormalize $pd$ scattering at NLO\cite{Vanasse:2014kxa}.  Thus an additional Coulomb three-body force, $H_{0,1}^{(\alpha)}(\Lambda)$,  is required for $pd$ scattering at NLO.  Fitting this new three-body force to the $\jjvHe$ binding energy and using the ERE at NLO gives the results in Fig.~\ref{fig:H1a}. 
In the ERE, $\rho_{s}$ is the effective range for $np$ scattering in the ${}^{1}\!S_{0}$ channel and $r_{C}$ is the effective range for $pp$ scattering.  Note that these experimentally are very close to each other, but are different due to isospin breaking effects.  When $\rho_{s}\neq r_{C}$, $H_{0,1}^{(\alpha)}(\Lambda)$ is dominated by a linear divergence.  To remove this linear divergence and convert poles to zeroes Ref.~\refcite{Vanasse:2014kxa} plotted $\Lambda/H_{0,1}^{(\alpha)}(\Lambda)$, shown in Fig.~\ref{subfig:H1aNot}.  For the case $\rho_{s}=r_{C}$, $H_{0,1}^{(\alpha)}(\Lambda)$ is dominated by a $\log(\Lambda)^{2}$ divergence.  This is shown Fig.~\ref{subfig:H1aIs} where $\log(\Lambda)^{2}/H_{0,1}^{(\alpha)}(\Lambda)$ is plotted to divide out the dominant $\log(\Lambda)^{2}$ behavior and convert all poles to zeroes.  The numerical calculations given by black dots match the analytical predictions shown with red lines well, above $\Lambda=5000$~MeV.  Above this cutoff the $pd$ scattering amplitude is well approximated by its analytically determined asymptotic form and thus shows good agreement with numerical results.  

\begin{figure}[th]
\centerline{\includegraphics[width=80mm]{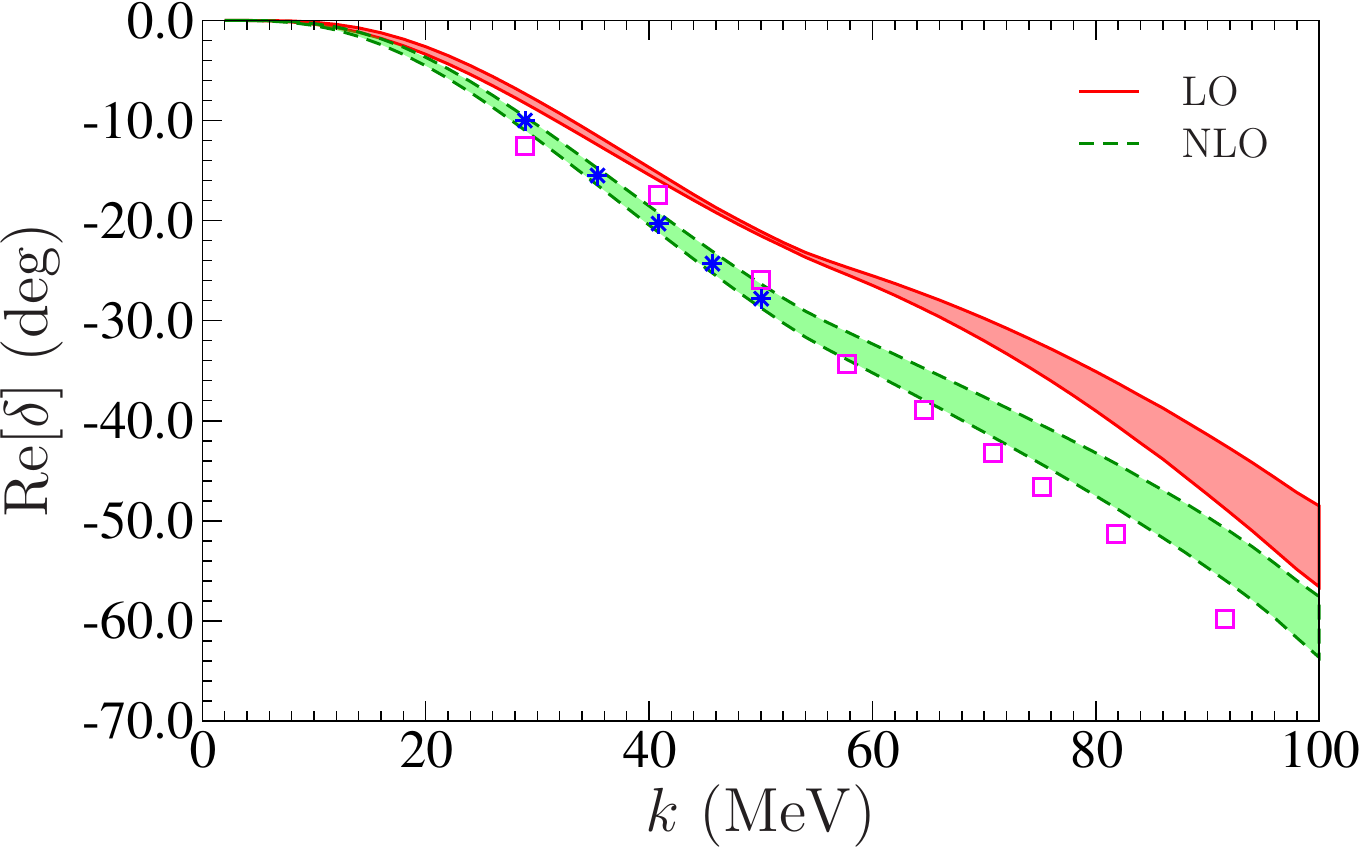}}
\caption{\label{fig:pdPhase} Real part of Coulomb subtracted phase shift for $pd$ scattering in the doublet $S$-wave channel.  The red band with solid line borders corresponds to the LO calculation and the green band with dashed line borders the NLO calculation.  The width of the bands corresponds to cutoff variation in which the cutoff is varied from $\Lambda$=200-10$^{7}$~MeV.  The blue stars are calculations using AV-18+UR with a hyperspherical harmonics approach\cite{Kievsky:1996ca}, and the pink squares are a phase shift analysis from experimental data\cite{Arvieux:1974} (Figure from Ref.~\citen{Vanasse:2014kxa}).}
\end{figure}

Using the new three-body force $H_{0,1}^{(\alpha)}(\Lambda)$ Ref.~\refcite{Vanasse:2014kxa} then calculated the $pd$ doublet $S$-wave Coulomb subtracted phase shift up to NLO shown in Fig.~\ref{fig:pdPhase}.
The variation of the bands corresponds to varying the cutoff from $\Lambda=200-10^{7}$~MeV, and convergence for large cutoffs was observed.  The red band with solid borders is the LO prediction and the green band with dashed borders the NLO prediction.  The blue stars are from PMC using AV18+UR and a hyperspherical harmonics approach \cite{Kievsky:1996ca}.  Finally the pink squares come from a phase shift analysis of experimental data\cite{Arvieux:1974}.  The NLO \EFT calculation agrees with the available data and PMC up to the expected 17\% error ($[\gamma_{t}\rho_{t}]^{2}\approx.17$) in the ERE, where $\rho_{t}=1.765$~fm is the effective range about the deuteron pole.

Further numerical evidence for the need of a NLO Coulomb three-body force was given by K{\"o}nig et al\cite{Konig:2014ufa} using the partial-resummation technique.  It should be noted that two three-body force terms are not required for $pd$ scattering at NLO.  If a description of only $pd$ properties at NLO is desired, only one such force is required.  Rather the conclusion is that the same three-body force cannot be used for both $nd$ and $pd$ at NLO\cite{Vanasse:2014kxa,Konig:2014ufa}.  Recently K{\"o}nig et al.~showed that if the Coulomb interaction is treated perturbatively, $\rho_{s}=r_{C}$, and the the spin singlet channel is expanded perturbatively about the unitary limit that only one three-body force is needed for $nd$ and $pd$ scattering at NLO~\cite{Konig:2015aka}.  Finally, using strictly perturbative techniques and Coulomb corrections, Hammer and K{\"o}nig predicted the quartet $S$-wave channel $pd$ scattering length\cite{Konig:2013cia}.

\section{\label{sec:3BBreakup} Three-Body Breakup}

\begin{figure}[th]
\centerline{\includegraphics[width=100mm]{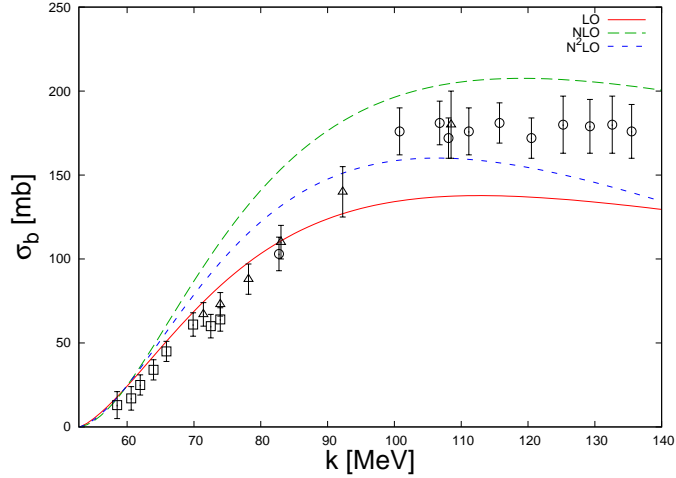}}
\caption{\label{fig:3BBreakup}Momentum dependence in the c.m.~frame of the three-body breakup cross-section in $n+d\to n+n+p$.  The solid red line, long-dashed green line, and the short dashed blue line are the LO, NLO, and N$^{2}$LO predictions in \EFT respectively.  The open squares are data from Holmberg and Hans{\'e}n, the open triangles data from Catron et al., and the open circles data from Pauletta and Brooks \cite{Holmberg1969305,PhysRev.123.218,Pauletta1975267}. }
\end{figure}

The three-body breakup process $n+d\to n+n+p$ can also be treated in \EFT.  This process has been calculated using separable potentials in a calculation very similar to a \EFT calculation\cite{Aaron:1966zz}.  Rather than calculating the three-body breakup amplitude $n+d\to n+n+p$ the total three-body breakup cross-section can be related to the $nd$ scattering amplitude via unitarity\cite{Kievsky:2000eb}, which yields\footnote{Note that there is also the open channel $nd\to t\gamma$.  However, this channel is suppressed by factors of $\alpha_{em}$ and is comparatively small at three-body breakup energies and can therefore be ignored at this order in the unitarity argument.  The calculation does not posses any photons, so the unitarity argument is rigorous in the calculation but not in the physical process.}
\begin{equation}
\sigma_{b}=\frac{\pi}{k^{2}}\frac{1}{6}\sum_{J}(2J+1)\frac{2kM_{N}}{3\pi}\sum_{\alpha}\left[2\mathrm{Im}\left[T^{J}_{\alpha,\alpha}\right]-\frac{2kM_{N}}{3\pi}\sum_{\beta}|T^{J}_{\alpha,\beta}|^{2}\right]
\end{equation}
where $T^{J}_{\alpha,\beta}$ are the $nd$ scattering amplitudes, and $\alpha=L',S'$, and $\beta=L,S$.  Here $L$ ($S$) is the relative orbital angular momentum (total spin angular momentum) magnitude in $nd$ scattering, $J$ the total angular momentum magnitude, and $k$ the c.m.~momentum.  This gives the results in Fig.~\ref{fig:3BBreakup} for the total three-body breakup cross-section.
The solid red line is the LO prediction, the long-dashed green line the NLO prediction, and the short-dashed blue line the N$^{2}$LO prediction.  The open squares are data from Holmberg and Hans{\'e}n, the open triangles data from Catron et al., and the open circles data from Pauletta and Brooks \cite{Holmberg1969305,PhysRev.123.218,Pauletta1975267}.  At momenta below 70 MeV the LO, NLO, and N$^{2}$LO curves seem to slightly overpredict the experimental data.  The poorer prediction of the data at these momenta should not come as a surprise, since here the $P$-wave contributions become important and the $P$-wave phase shifts in comparison to PMC are reproduced worse at NLO and N$^{2}$LO than LO at these momenta\cite{Griesshammer:2004pe,Vanasse:2013sda}.  As the momenta is increased above 70~MeV we see further disagreement with the data.  However, at these momenta the effective theory breaks down since the momentum breakdown scale is roughly $m_{\pi}/2$, the momentum at which the t-channel cut from potential pion exchange occurs.  It seems the use of \EFT in the three-body breakup channel has limited use as it can only describe a small window of energies.  Therefore, it will be important to develop a consistent pionful theory to properly investigate three-body breakup observables.

\section{\label{sec:conclusion}Conclusions}

In comparison with PMC and available data \EFT has been very successful in reproducing phase shifts in both $nd$ and $pd$ scattering\cite{Gabbiani:1999yv,Griesshammer:2004pe,Vanasse:2013sda,Konig:2011yq}.  The newer perturbative techniques qualitatively agree with the earlier partial resummation technique.  However, the perturbative techniques give the correct sign for the imaginary part of the quartet $S$-wave phase shift above the deuteron breakup threshold.  In the strictly perturbative scheme the $SD$ mixing term was included at N$^{2}$LO and good agreement was found with many of the eigen-phases and mixing angles in comparison to PMC\cite{Vanasse:2013sda}.  Significant differences were found for the $\epsilon_{J}^{\pi}$ mixing angles, as well as the quartet $P$-wave eigen-phases\footnote{The $\epsilon_{J}^{\pi}$ mixing angles, mix partial waves of different $S$ values but the same $L$ value}.  A recent strictly perturbative N$^{3}$LO calculation of $nd$ scattering including the contributions from two-body $P$-wave contact interactions has been performed~\cite{Margaryan:2015rzg} and found the $A_{y}$ polarization observables were especially sensitive to the values of the two-body $P$-wave contact interactions\footnote{Note PMC find a similar sensitivity to the two-body $P$-wave channels~\cite{Witala1989446,WITALA199148}.}.  As a result, the $A_{y}$ observables naively have a large error associated with them making them consistent with data at N$^{3}$LO.  A higher order calculation will be necessary to reduce these errors and make a better comparison with available experimental data.

The capture reaction $nd\to t\gamma$ has been calculated at NNLO in the PC sector and at LO in the PV sector\cite{Arani:2011if,Arani:2014qsa}.  Use of the new perturbative techniques offer a simple approach to calculate higher order contributions.  In general the new perturbative technique can be used to add higher order corrections to any diagram with external currents.  This allows a description of processes at low energies such as $\gamma+t\to\gamma+t$,$\gamma+t\to n+d$,$\gamma+\jjvHe\to\gamma+\jjvHe$,$\gamma+\jjvHe\to p+d$, and $t\to\jjvHe+e^{-}+\bar{\nu}_{e}$ in both PC and PV sectors.  However, it seems \EFT will not offer much information in three-body breakup processes such as $\gamma+t\to n+n+p$ due to the limited energy range available to \EFT.

In the bound state regime much more work is needed in \EFT.  The new perturbative techniques reviewed here will be of great utility in this endeavor as they provide a straightforward inclusion of perturbative corrections to bound state calculations.  These methods give a LO triton charge radius of 1.13~fm and a NLO triton charge radius of 1.59~fm in agreement with the experimental result of 1.755$\pm$.086~fm, within the expected error of \EFT\cite{Vanasse:2015fph} at NLO.  The LO result is more than 30\% away from the experimental value, which is greater than the naive LO error estimate of \EFT.  A different LO calculation of this quantity in \EFT using a wavefunction approach gives a value of 2.1$\pm$.6~fm, which agrees with the experimental result within the expected LO error of \EFT\cite{Platter:2005sj}.  These new perturbative techniques also allow for a more efficient calculation of three-body forces and perturbative corrections to binding energies.  

All of the strictly perturbative techniques described here are equally useful in the description of $pd$ scattering and bound state properties of $\jjvHe$, but the additional Coulomb interaction will have to be taken into account.  Complications due to the Coulomb interaction have been shown at NLO in $pd$ scattering where the same NLO three-body force cannot be used for $nd$ and $pd$ scattering\cite{Vanasse:2014kxa}.  Thus for a consistent picture of both $nd$ and $pd$ scattering a new isospin-dependent three-body force is required.  This means that at NLO and likely higher orders $nd$ data alone cannot be used to renormalize counter-terms in $pd$ scattering.  At N$^{2}$LO in $pd$ scattering two renormalization conditions requiring $pd$ data will likely be needed.  In $nd$ scattering the usual renormalization conditions are the $nd$ scattering length and the $\jjvH$ binding energy, but fitting to the $pd$ scattering length will be complicated due to Coulomb interactions.  With the new perturbative method using bound state properties of $\jjvHe$ is feasible.  Further work is required to consider the most efficient renormalization conditions for $pd$ scattering at higher orders. Calculating $pd$ scattering to higher order is desirable to investigate polarization observables since there is much more experimental data for polarization observables in $pd$ than $nd$ scattering.

Finally, the use of \EFT for three-body breakup observables was discussed.  Using unitarity the three-body breakup cross-section can be related to the $nd$ scattering amplitudes.  Comparing the \EFT predictions to the available breakup data for $nd$  gave a general over-prediction of the data.  The momentum breakdown scale of \EFT occurs at $\Lambda_{\not{\pi}}=m_{\pi}/2$, the momentum at which the t-channel cut from potential pion exchange occurs.  It is likely the disagreement with data is due to the fact that above the deuteron breakup threshold the breakdown scale of \EFT is quickly approached.  Thus it seems the use of \EFT in three-body breakup is very limited, only a small window between 50-70 MeV exists where it is strictly valid, and even at these higher momenta signs of breakdown may already be visible. As a result it is important that a pionful theory that is renormalization group invariant be developed.  Such a theory will allow for the calculation of three-body breakup observables such as the symmetric space star and quasi-free scattering configurations of outgoing particles for which there currently exists discrepancies between theory and experiment\cite{PhysRevC.71.034006}.

\section*{Acknowledgements}
I would like to thank Thomas Mehen and Roxanne Springer for useful conversations during the course of this work.  I am also appreciative for comments on the manuscript from Roxanne Springer.  This work is supported in part by the US Department of Energy under Grant no. DE-FG02-05ER41368.


\begin{thebibliography}{10}

\bibitem{Bedaque:1997qi}
P.~F. Bedaque and U.~van Kolck, {\em Phys. Lett.} {\bf B428}  (1998) 221.

\bibitem{vanKolck:1997ut}
U.~van Kolck, {\em Lect. Notes Phys.} {\bf 513} (1998) 62.

\bibitem{Kaplan:1998tg}
D.~B. Kaplan, M.~J. Savage and M.~B. Wise, {\em Phys. Lett. B} {\bf 424}
  (1998) 390.

\bibitem{Kaplan:1998we}
D.~B. Kaplan, M.~J. Savage and M.~B. Wise, {\em Nucl. Phys. B} {\bf 534}
  (1998) 329.

\bibitem{Chen:1999tn}
J.-W. Chen, G.~Rupak and M.~J. Savage, {\em Nucl. Phys. A} {\bf 653}  (1999)
  386.

\bibitem{Ando:2004mm}
S.-i. Ando and C.~H. Hyun, {\em Phys. Rev. C} {\bf 72}  (2005)   014008.

\bibitem{Ando:2007fh}
S.-i. Ando, J.~W. Shin, C.~H. Hyun and S.~W. Hong, {\em Phys. Rev. C} {\bf 76}
  (2007)   064001.

\bibitem{Kong:1999sf}
X.~Kong and F.~Ravndal, {\em Nucl. Phys. A} {\bf 665}  (2000) 137.

\bibitem{Chen:1999bg}
J.-W. Chen and M.~J. Savage, {\em Phys. Rev. C} {\bf 60}  (1999)   065205.

\bibitem{Rupak:1999rk}
G.~Rupak, {\em Nucl. Phys. A} {\bf 678}  (2000) 405.

\bibitem{Ando:2005cz}
S.~Ando, R.~Cyburt, S.~Hong and C.~Hyun, {\em Phys. Rev. C} {\bf 74}  (2006)
  025809.

\bibitem{Schindler:2009wd}
M.~R. Schindler and R.~P. Springer, {\em Nucl. Phys. A} {\bf 846}  (2010) 51.

\bibitem{Phillips:2008hn}
D.~R. Phillips, M.~R. Schindler and R.~P. Springer, {\em Nucl. Phys. A} {\bf
  822}  (2009) 1.

\bibitem{Shin:2009hi}
J.~Shin, S.~Ando and C.~Hyun, {\em Phys. Rev. C} {\bf 81}  (2010)   055501.

\bibitem{Vanasse:2014sva}
J.~Vanasse and M.~R. Schindler, {\em Phys. Rev.} {\bf C90}  (2014)   044001.

\bibitem{Kong:2000px}
X.~Kong and F.~Ravndal, {\em Phys. Rev. C} {\bf 64}  (2001)   044002.

\bibitem{Butler:2000zp}
M.~Butler, J.-W. Chen and X.~Kong, {\em Phys. Rev. C} {\bf 63}  (2001)
  035501.

\bibitem{Ando:2008va}
S.~Ando, J.~Shin, C.~Hyun, S.~Hong and K.~Kubodera, {\em Phys. Lett. B} {\bf
  668}  (2008) 187.

\bibitem{Chen:2012hm}
J.-W. Chen, C.-P. Liu and S.-H. Yu, {\em Phys. Lett. B} {\bf 720}  (2013) 385.

\bibitem{Kirscher:2009aj}
J.~Kirscher, H.~W. Grie{\ss}hammer, D.~Shukla and H.~M. Hofmann, {\em Eur. Phys.
  J.} {\bf A44}  (2010) 239.

\bibitem{Kirscher:2011uc}
J.~Kirscher, {\em Phys. Lett.} {\bf B721}  (2013) 335.

\bibitem{Kirscher:2015ana}
J.~Kirscher, {Pionless Effective Field Theory in Few-Nucleon Systems}, PhD
  thesis  (2015).

\bibitem{Phillips:1999hh}
D.~R. Phillips, G.~Rupak and M.~J. Savage, {\em Phys. Lett. B} {\bf 473}
  (2000) 209.

\bibitem{Griesshammer:2004pe}
H.~W. Grie{\ss}hammer, {\em Nucl. Phys. A} {\bf 744}  (2004) 192.

\bibitem{Bedaque:1998mb}
P.~F. Bedaque, H.-W.~Hammer and U.~van Kolck, {\em Phys. Rev. C} {\bf 58}  (1998)
  641.

\bibitem{Bedaque:1998kg}
P.~F. Bedaque, H.-W.~Hammer and U.~van Kolck, {\em Phys. Rev. Lett.} {\bf 82}
  (1999) 463.

\bibitem{Bedaque:1998km}
P.~F. Bedaque, H.-W.~Hammer and U.~van Kolck, {\em Nucl. Phys. A} {\bf 646}
  (1999) 444.

\bibitem{Bedaque:1999ve}
P.~F. Bedaque, H.-W.~Hammer and U.~van Kolck, {\em Nucl. Phys. A} {\bf 676}
  (2000) 357.

\bibitem{Hammer:2000nf}
H.-W. Hammer and T.~Mehen, {\em Nucl. Phys.} {\bf A690}  (2001) 535.

\bibitem{Bedaque:1999vb}
P.~F. Bedaque and H.~W. Grie{\ss}hammer, {\em Nucl. Phys. A} {\bf 671}  (2000)
  357.

\bibitem{Hammer:2001gh}
H.-W.~Hammer and T.~Mehen, {\em Phys. Lett. B} {\bf 516}  (2001) 353.

\bibitem{Gabbiani:1999yv}
F.~Gabbiani, P.~F. Bedaque and H.~W. Grie{\ss}hammer, {\em Nucl. Phys. A} {\bf
  675}  (2000) 601.

\bibitem{Bedaque:2002yg}
P.~F. Bedaque, G.~Rupak, H.~W. Grie{\ss}hammer and H.-W. Hammer, {\em Nucl.
  Phys. A} {\bf 714}  (2003) 589.

\bibitem{Platter:2006ev}
L.~Platter and D.~R. Phillips, {\em Few Body Syst.} {\bf 40}  (2006) 35.

\bibitem{Ji:2012nj}
C.~Ji and D.~R. Phillips, {\em Few-Body Syst.} {\bf 54}  (2013) 2317.

\bibitem{Gabbiani:2001yh}
F.~Gabbiani  (2001), nucl-th/0104088.

\bibitem{Griesshammer:2005ga}
H.~W. Grie{\ss}hammer, {\em Nucl. Phys. A} {\bf 760}  (2005) 110.

\bibitem{Birse:2005pm}
M.~C. Birse, {\em J. Phys.} {\bf A39}  (2006)   L49.

\bibitem{Griesshammer:2010nd}
H.~W. Grie{\ss}hammer and M.~R. Schindler, {\em Eur. Phys. J. A} {\bf 46}
  (2010) 73.

\bibitem{Vanasse:2011nd}
J.~Vanasse, {\em Phys. Rev. C} {\bf 86}  (2012)   014001.

\bibitem{Griesshammer:2011md}
H.~W. Grie{\ss}hammer, M.~R. Schindler and R.~P. Springer, {\em Eur. Phys. J.
  A} {\bf 48}  (2012)  ~7.

\bibitem{Vanasse:2013sda}
J.~Vanasse, {\em Phys. Rev. C} {\bf 88}  (2013)   044001.

\bibitem{Vanasse:2015fph}
J.~Vanasse  (2015), 1512.03805.

\bibitem{Hagen:2013xga}
P.~Hagen, H.-W. Hammer and L.~Platter, {\em Eur.Phys.J.} {\bf A49}  (2013)
  118.

\bibitem{Hammer:2014rba}
H.-W. Hammer and S.~K{\"o}nig, {\em Phys. Lett.} {\bf B736}  (2014) 208.

\bibitem{Margaryan:2015rzg}
A.~Margaryan, R.~P. Springer and J.~Vanasse  (2015), 1512.03774.

\bibitem{Skornyakov}
G.V.Skornyakov and K.A.Ter-Martirosian, {\em Sov. Phys. JETP} {\bf 4}  (1957)
  648.

\bibitem{press1996numerical}
W.~H. Press, S.~A. Teukolsky, W.~T. Vetterling and B.~P. Flannery, {\em
  Numerical recipes in C} (Cambridge university press Cambridge, 1996).

\bibitem{delves1988computational}
L.~M. Delves and J.~Mohamed, {\em Computational methods for integral equations}
  (CUP Archive, 1988).

\bibitem{glockle2012quantum}
W.~Gl{\"o}ckle, {\em The quantum mechanical few-body problem} (Springer Science
  \& Business Media, 2012).

\bibitem{Hetherington:1965zza}
J.~H. Hetherington and L.~H. Schick, {\em Phys. Rev.} {\bf 137}  (1965) B935.

\bibitem{Ziegelmann}
E.~Schmid and H.~Ziegelmann, {\em The Qauntum Mechanical Three-Body Problem,
  Vieweg Tract in Pure and Applied Physics Vol. 2} (Pergamon Press, 1974).

\bibitem{Brayshaw:1969ab}
D.~D. Brayshaw, {\em Phys. Rev.} {\bf 176}  (1968) 1855.

\bibitem{Konig:2011yq}
S.~K{\"o}nig and H.-W. Hammer, {\em Phys. Rev. C} {\bf 83}  (2011)   064001.

\bibitem{Mehen:1999qs}
T.~Mehen, I.~W. Stewart and M.~B. Wise, {\em Phys.Rev.Lett.} {\bf 83}  (1999)
  931.

\bibitem{PhysRev.51.106}
E.~Wigner, {\em Phys. Rev.} {\bf 51} (Jan 1937) 106.

\bibitem{Platter:2005sj}
L.~Platter and H.-W. Hammer, {\em Nucl. Phys.} {\bf A766}  (2006) 132.

\bibitem{Arani:2011if}
M.~Moeini~Arani and S.~Bayegan, {\em Eur. Phys. J.} {\bf A49}  (2013)   117.

\bibitem{Arani:2014qsa}
M.~M. Arani, H.~Nematollahi, N.~Mahboubi and S.~Bayegan, {\em Phys. Rev.} {\bf
  C89}  (2014)   064005.

\bibitem{Konig:2014ufa}
S.~K{\"o}nig, H.~W. Grie{\ss}hammer and H.-W. Hammer, {\em J. Phys.} {\bf G42}
  (2015)   045101.

\bibitem{Ando:2010wq}
S.-i. Ando and M.~C. Birse, {\em J. Phys. G} {\bf 37}  (2010)   105108.

\bibitem{Amroun:1994qj}
A.~Amroun {\em et~al.}, {\em Nucl. Phys.} {\bf A579}  (1994) 596.

\bibitem{Rupak:2001ci}
G.~Rupak and X.-w. Kong, {\em Nucl. Phys. A} {\bf 717}  (2003) 73.

\bibitem{Alt1985429}
E.~Alt, W.~Sandhas and H.~Ziegelmann, {\em Nuclear Physics A} {\bf 445}  (1985)
  429 .

\bibitem{levin2013coulomb}
F.~S. Levin and D.~A. Micha, {\em Coulomb interactions in nuclear and atomic
  few-body collisions} (Springer Science \& Business Media, 2013).

\bibitem{Vanasse:2014kxa}
J.~Vanasse, D.~A. Egolf, J.~Kerin, S.~K{\"o}nig and R.~P. Springer, {\em Phys.
  Rev.} {\bf C89}  (2014)   064003.

\bibitem{Kievsky:1996ca}
A.~Kievsky, S.~Rosati, W.~Tornow and M.~Viviani, {\em Nucl. Phys. A} {\bf 607}
  (1996) 402.

\bibitem{Arvieux:1974}
J.~Arvieux, {\em Nucl. Phys. A} {\bf 221}  (1974) 253.

\bibitem{Konig:2015aka}
S.~K{\"o}nig, H.~W. Grie{\ss}hammer, H.-W. Hammer and U.~van Kolck  (2015), 1508.05085.

\bibitem{Konig:2013cia}
S.~K{\"o}nig and H.-W. Hammer, {\em Phys. Rev.} {\bf C90}  (2014)   034005.

\bibitem{Aaron:1966zz}
R.~Aaron and R.~D. Amado, {\em Phys. Rev.} {\bf 150}  (1966) 857.

\bibitem{Kievsky:2000eb}
A.~Kievsky, C.~R. Brune and M.~Viviani, {\em Phys. Lett.} {\bf B480}  (2000)
  250.

\bibitem{Holmberg1969305}
M.~Holmberg and J.~Hans{\'e}n, {\em Nuclear Physics A} {\bf 129}  (1969) 305 .

\bibitem{PhysRev.123.218}
H.~C. Catron, M.~D. Goldberg, R.~W. Hill, J.~M. LeBlanc, J.~P. Stoering, C.~J.
  Taylor and M.~A. Williamson, {\em Phys. Rev.} {\bf 123} (Jul 1961) 218.

\bibitem{Pauletta1975267}
G.~Pauletta and F.~Brooks, {\em Nuclear Physics A} {\bf 255}  (1975) 267 .

\bibitem{Witala1989446}
H.~Wita{\l}a, W.~Gl{\"o}ckle and T.~Cornelius, {\em Nuclear Physics A} {\bf 496}
  (1989) 446 .

\bibitem{WITALA199148}
H.~Wita{\l}a and W.~Gl{\"o}ckle, {\em Nuclear Physics A} {\bf 528}  (1991) 48 .

\bibitem{PhysRevC.71.034006}
H.~R. Setze, C.~R. Howell, W.~Tornow, R.~T. Braun, D.~E. Gonz\'alez~Trotter,
  A.~H. Hussein, R.~S. Pedroni, C.~D. Roper, F.~Salinas,
  I.~\ifmmode~\check{S}\else \v{S}\fi{}laus, B.~Vlahovi\ifmmode~\acute{c}\else
  \'{c}\fi{}, R.~L. Walter, G.~Mertens, J.~M. Lambert, H.~Wita\l{}a and
  W.~Gl\"ockle, {\em Phys. Rev. C} {\bf 71} (Mar 2005)   034006.

\end{thebibliography}

\end{document}